\documentclass[transmag]{IEEEtran}
\usepackage{latexsym}
\usepackage{graphicx}
\usepackage{amsfonts,amssymb,amsmath}
\usepackage{hyperref}

\usepackage{epsfig}
\usepackage{algorithm}
\usepackage{algorithmic}
\usepackage{amsmath}
\usepackage{amssymb}
\usepackage{color}
\usepackage{url}
\usepackage{comment}
\usepackage{subcaption}
\usepackage{multirow}
\usepackage{tikz}
\usepackage[export]{adjustbox}
\usepackage{array}

\def\BibTeX{{\rm B\kern-.05em{\sc i\kern-.025em b}\kern-.08em T\kern-.1667em\lower.7ex\hbox{E}\kern-.125emX}}

\begin{document}

\title{ANFIC: Image Compression Using Augmented Normalizing Flows}

\author{Yung-Han Ho, Chih-Chun Chan, Wen-Hsiao Peng, \IEEEmembership{Senior Member, IEEE}, Hsueh-Ming Hang, \IEEEmembership{Fellow, IEEE}, \\and Marek Domański, \IEEEmembership{Senior Member, IEEE}
%\IEEEmembership{Member, IEEE}

%\thanks{This paragraph of the first footnote will contain the date on which you submitted your paper for review. It will also contain support information, including sponsor and financial support acknowledgment. For example, ``This work was supported in part by the U.S. Department of Commerce under Grant BS123456''.}

\thanks{Manuscript received July 1, 2021; revised September 29, 2021; accepted October 15, 2021. This work was supported by National Center for High-Performance Computing, Taiwan.}

\thanks{Yung-Han Ho, Chih-Chun Chan, and Wen-Hsiao Peng are with the Department of Computer Science, National Yang Ming Chiao Tung University, Hsinchu, Taiwan (e-mail: wpeng@cs.nctu.edu.tw).}

\thanks{Hsueh-Ming Hang is with the Department of Electronics Engineering, National Yang Ming Chiao Tung University, Hsinchu, Taiwan (e-mail: hmhang@nctu.edu.tw).}

\thanks{Marek Domański is with the Institute of Multimedia Telecommunications, Poznań University of Technology, Poznań, Poland (e-mail: marek.domanski@put.poznan.pl).}

%\thanks{This work is supported by Qualcomm technologies, Inc. (NAT-439543) and National Center for High-Performance Computing, Taiwan.}
}

\IEEEtitleabstractindextext{\begin{abstract}
This paper introduces an end-to-end learned image compression system, termed ANFIC, based on Augmented Normalizing Flows (ANF). ANF is a new type of flow model, which stacks multiple variational autoencoders (VAE) for greater model expressiveness. The VAE-based image compression has gone mainstream, showing promising compression performance. Our work presents the first attempt to leverage VAE-based compression in a flow-based framework. ANFIC advances further compression efficiency by stacking and extending hierarchically multiple VAE's. The invertibility of ANF, together with our training strategies, enables ANFIC to support a wide range of quality levels without changing the encoding and decoding networks. Extensive experimental results show that in terms of PSNR-RGB, ANFIC performs comparably to or better than the state-of-the-art learned image compression. Moreover, it performs close to VVC intra coding, from low-rate compression up to perceptually lossless compression. In particular, ANFIC achieves the state-of-the-art performance, when extended with conditional convolution for variable rate compression with a single model. The source code of ANFIC can be found at \url{https://github.com/dororojames/ANFIC}.

%utilizes invertible and bijective coupling layers to losslessly map the image into the latent space. Inspired from ANF, which is composed of multiple variational autoencoders (VAE) and is regarded as a generalization of VAE. Our ANFIC provides a greater transformation ability by extending the existing VAE-based methods into an ANF-based image compression. Deeper into the mechanism of ANFIC, each layer of the ANF operates as a low-pass filter to remove the high-frequency information from the input image and yielding a nearly zero low-frequency residual signal in the end. Experimental results show that our ANFIC achieves competitive performance compared with the state-of-the-art for both PSNR-RGB and MS-SSIM on three common test datasets. Due to the inherent property of normalizing flows, our ANFIC operates in a wide range across low to high rates. In addition, We investigate the effectiveness of our ANFIC compared with the VAE-based scheme, discuss the impact of the quality enhancement network to reduce the quantization error, and accomplish the unified variable bitrate model without the performance degradation. 
\end{abstract}

\begin{IEEEkeywords}
Learning-based image compression, flow-based image compression, augmented normalizing flows, perceptually lossless image compression, variable rate image compression
%At least four keywords or phrases in 
%alphabetical order, separated by commas. For a list of suggested keywords, 
%send a blank e-mail to %\href{mailto:keywords@ieee.org}{mailto:keywords@ieee.org} or visit 
%\href{http://www.ieee.org/organizations/pubs/ani_prod/keywrd98.txt}{http://www.ieee.org/organizations/pubs/ani\_prod/keywrd98.txt}
\end{IEEEkeywords}

%Note: There should no nonstandard abbreviations, acknowledgments of support, 
%references or footnotes in in the abstract.
}

\maketitle

\section{Introduction}\label{sec:introduction}
Image compression has been a thriving research area for decades due to the storage and transmission requirements in various applications that underpin our modern digital life. Image compression also appears in the form of intra-frame coding for video compression~\cite{dvc}. The rapid advances in inter-frame prediction make efficient intra-frame coding become increasingly important because intra-coded frames often predominate over the bit rate of a compressed video. Therefore, it is much desirable to achieve even higher image compression efficiency. 

The state-of-the-art image compression methods, e.g. BPG and VVC intra coding, usually involve block-based intra prediction, block-based transform coding of residuals, and context-adaptive binary arithmetic coding. Over the years, tremendous research effort has been invested to better every component in a way that seeks higher compression efficiency at the expense of an acceptable complexity increase. These hand-crafted codecs, although achieving a good balance between compression efficiency and complexity, lacks the opportunity to optimize all the components jointly in a seamless, end-to-end manner.

The rising of deep learning recently spurred a new wave of developments in image compression, with end-to-end learned systems attracting lots of attention. Among them, the variational autoencoder (VAE)-based methods~\cite{ContextModel,coarse,attn,ContextGMM} have achieved compression performance very close to the latest VVC intra coding. Different from traditional hand-crafted codecs, the VAE-based methods usually implement an image-level non-linear transform that converts an input image into a compact set of latent features, the dimensions of which are much smaller than the input image. Ever since the advent of the first VAE-based scheme~\cite{GoogleFactorized}, several improvements have been made on the expressiveness~\cite{attn,ContextGMM,recurrent} of the autoencoder and the efficiency of entropy coding~\cite{ContextModel,coarse,attn,ContextGMM,GoogleHyperPrior,lee,benchmark}. Up to now, the VAE-based methods have become the mainstream approach to end-to-end learned image compression. 

However, one issue with most VAE-based schemes is that the autoencoder is generally lossy. There is no guarantee that its non-linear transform can reconstruct the input image losslessly even without quantizing the latent features of the image. This is unlike the traditional transforms, such as Discrete Cosine Transform and Wavelet Transform, which have the desirable property of perfect reconstruction and allow the codec to offer a wide range of quality levels by merely changing the quantization step size.

Recently, the flow-based models~\cite{NFcodec,iwave} emerged as attractive alternatives. These models have the striking feature of realizing a bijective and invertible mapping between the input image and its latent features via the use of reversible networks composed of affine coupling layers~\cite{NF,nvp}. This invertibility is utilized to develop lossless image compression in~\cite{IDF}, while the affine coupling layers are used in place of the lossy autoencoder in~\cite{NFcodec,iwave} to achieve both lossy and lossless (or perceptually lossless) compression with a single unified model. The reversible networks, however, are quite distinct from the commonly used autoencoders, making these two types of compression systems not compatible with each other.     

In this paper, we propose a novel end-to-end lossy image compression system, termed ANFIC, based on Augmented Normalizing Flows (ANF)~\cite{ANF}. ANF is a new type of flow models that work on augmented input space to offer greater transformation ability than the ordinary flow models. Our scheme ANFIC is motivated by the fact that ANF is a generalization of VAE that stacks multiple VAE's as a flow model. In a sense, this allows ANFIC to extend any existing VAE-based compression system in a flow-based framework to enjoy the benefits of both approaches. ANFIC is novel and unique in that (1) it distinguishes from flow-based compression by operating in augmented input space, being able to leverage the  representation power of any VAE-based image compression, and that (2) it is more general than the VAE-based compression by allowing VAE to be stacked and/or extended hierarchically. 

Extensive experimental results on Kodak, Tecnick, and CLIC validation datasets show that ANFIC performs comparably to or better than the state-of-the-art end-to-end image compression in terms of PSNR-RGB. It performs close to VVC intra over a wide range of quality levels from low-rate compression up to perceptually lossless compression. In particular, ANFIC achieves the state-of-the-art performance among the competing methods, when extended with conditional convolutional layers~\cite{conditional} for variate rate compression with a single model. 

Our main contributions are three-fold: 
\begin{itemize}
    \item We propose ANFIC, which uses augmented normalizing flows for image compression, as the first work that leverages VAE-based image compression in a flow-based framework.
    \item We offer extensive ablation studies to understand and visualize the inner workings of ANFIC.
    \item Extensive experimental results show that ANFIC is competitive with the state-of-the-art image compression, VAE-based and flow-based, over a wide range of quality levels and performs close to VVC intra coding.
    \end{itemize}
This work improves on our previous publication~\cite{Ho} by (1) replacing the affine coupling layers with additive coupling layers to improve the training stability and avoid degrading the performance, (2) introducing the Gaussian mixture model along with the autoregressive module for better entropy coding, and (3) providing more comprehensive ablation studies of ANFIC.   

The remainder of this paper is organized as follows: Section~\ref{sec:related} reviews VAE-based image compression and the basics of ANF. Section~\ref{sec:method} elaborates the design of ANFIC. Section~\ref{sec:experiment} compares ANFIC with the state-of-the-art methods in terms of objective compression performance and subjective image quality. Section~\ref{sec:ablation} presents our ablation studies. Finally, we provide concluding remarks in Section~\ref{sec:conclude}.

\section{Related Work} 
\label{sec:related}
In this paper, we propose an ANF-based image compression. It can be viewed as an extension of VAE-based image compression. Hence, this section focuses on the recent developments of VAE-based image compression and introduces the fundamentals of ANF to ease the understanding of our scheme.

\begin{comment}
\begin{figure*}[t]
\begin{center}
\begin{subfigure}{0.24\textwidth}
    % include first image
    \centering
    %\vspace{-0.8em}
    \includegraphics[width=\linewidth]{FIG/factorize.png} 
    \caption{}
    %\vspace{-1.2em}
    \label{fig:Factorized}
\end{subfigure}
\begin{subfigure}{0.24\textwidth}
    % include first image
    \centering
    %\vspace{-0.8em}
    \includegraphics[width=\linewidth]{FIG/hyper.png}
    \caption{}
    %\vspace{-1.2em}
    \label{fig:Hyper}
\end{subfigure}
\begin{subfigure}{0.24\textwidth}
    % include first image
    \centering
    %\vspace{-0.8em}
    \includegraphics[width=\linewidth]{FIG/gmm.png}  
    \caption{}
    %\vspace{-1.2em}
    \label{fig:Autoregressive}
\end{subfigure}
\begin{subfigure}{0.24\textwidth}
    % include first image
    \centering
    %\vspace{-0.8em}
    \includegraphics[width=\linewidth]{FIG/anf.png}  
    \caption{}
    %\vspace{-1.2em}
    \label{fig:ANFIC}
\end{subfigure}

\caption{VAE-based image compression with (a) factorized prior model, (b) hyperprior model, and (c) autoregressive Gaussian mixture model (GMM); (d) Our ANFIC with autoregressive GMM.}
\label{fig:models}
\end{center}
\vspace{-2.0em}
\end{figure*}
\end{comment}

\subsection{VAE-based Image Compression}
VAE-based image compression \cite{ContextModel,coarse,attn,ContextGMM,GoogleFactorized,GoogleHyperPrior,lee} is the most popular approach to end-to-end learned image compression. Its training framework includes three major components: the analysis transform, the prior distribution, and the synthesis transform. These components are implemented by neural networks. %The coding process of the vanilla VAE-based image compression can be formulated as:
%\begin{align}
%    &  y = g_a(x;\phi)\\
%    &  \hat{y} = Q(y)\\
%    &  \hat{x} = g_s(\hat{y};\theta),
%\label{eq:enc_qt_dec}
%\end{align}
%where $g_a$ is the analysis transform parameterized by $\phi$, $g_s$ is the synthesis transform parameterized by $\theta$, and $Q$ is the quantization. The $x, y, \hat{y}, \hat{x}$ stand for the original image, latent representation, quantized latent representation, and reconstructed image, respectively.

%As illustrated in Fig.~\ref{fig:Factorized}, 
The analysis transform $g_a$ encodes the raw image $x$ through an encoding distribution $q^{g_a}_\phi(y|x)$ with the latent representation $y$ uniformly quantized as $\hat{y}$. The $\hat{y}$ is then entropy encoded into a bitstream using a learned prior $p_\pi(\hat{y})$ implemented by a network $\pi$. Finally, the synthesis transform $g_s$ reconstructs approximately the input $x$ from $\hat{y}$ by a decoding distribution $p^{g_s}_\theta(x|\hat{y})$. 

All the network parameters are trained end-to-end by minimizing   
\begin{equation}
%\begin{split}
    \mathcal{L(\phi,\theta,\pi)} = \underbrace{-E_{q^{g_a}_\phi(\hat{y}|x)}[\log p^{g_s}_\theta(x|\hat{y})]}_{D}                                 \underbrace{-E_{q^{g_a}_\phi(\hat{y}|x)}[\log p_\pi(\hat{y})]}_{R},
\label{eq:R_D}
%\end{split}
\end{equation}
where the first term, denoted by $D$, aims to minimize the negative log-likelihood of $x$ and the second term minimizes the rate $R$ needed for signaling $\hat{y}$. In particular, it is shown that minimizing Eq.~\eqref{eq:R_D} amounts to maximizing the evidence lower bound (ELBO) of a latent variable model~\cite{vae}, which is specified by $p_\pi(\hat{y})$ and $p^{g_s}_\theta(x|\hat{y})$, with $q^{g_a}_\phi(\hat{y}|x)$ taking a uniform distribution that models the effect of uniform quantization. In a more general setting, a hyper-parameter $\lambda$ is introduced to balance between $D$ and $R$, yielding $\mathcal{L}=\lambda D + R$.

\emph{Balle et al.}~\cite{GoogleFactorized} are the first to introduce the aforementioned VAE framework together with a learned factorized prior to image compression. In entropy coding the image latents, they assume the prior distribution $p_\pi(\hat{y})$ over $\hat{y}$ to be factorial and learn the distribution by the network $\pi$. Their analysis and synthesis transforms are composed of convolutional neural networks and the general division normalization (GDN) layers, which originate from \cite{GDN}. 

Even since the advent of the VAE-based compression framework, several efforts have been made to advance its coding efficiency. In particular, some~\cite{ContextModel,coarse,attn,ContextGMM,GoogleHyperPrior,lee,benchmark} improve the prior estimation for better entropy coding while others~\cite{attn,ContextGMM,recurrent} address the analysis and syntheses transforms (referred collectively to as the autoencoding transform). We summarize briefly these efforts as follows. 

%Therefore, later on, many related works follow the same VAE-based concept but try different ways to modify both entropy coding and transformation to achieve better performances.

%Since the learned prior estimates the latent prior distribution in the statistics of the whole dataset by using the fully factorized model, the rate tend to be overestimated due to the mismatch between the global prior distribution and the posterior distribution. 

%\subsection{Enhanced Prior Estimation}
\textbf{Enhanced Prior Estimation:} The prior distribution $p_\pi(\hat{y})$ crucially determines the number of bits (i.e. the rate) needed to signal the quantized image latents $\hat{y}$. Recognizing the suboptimality of the factorized prior $p_\pi(\hat{y})$, where feature samples in every channel of $\hat{y}$ are independently and identically distributed, \emph{Balle et al.}~\cite{GoogleHyperPrior} propose the notion of hyperprior to model every feature sample separately by a Gaussian distribution. To this end, additional side information $\hat{z}$ is extracted from the image latent $y$ and sent to the decoder, making the density estimation of $\hat{y}$ dependent on the input $x$. 
%In symbols, we have
%\begin{align}
%    &  z = h_a(y;\phi_h)\\
%    &  \hat{z} = Q(z)\\
%    &  \mu, \sigma^2 = h_s(\hat{z};\theta_h),
%\label{eq:enc_qt_dec}
%\end{align}
%where $h_a,h_s$ are the hyper-analysis and hyper-synthesis transforms parameterized by $\phi_h,\theta_h$; $Q$ is the quantization, $\mu$ and $\sigma^2$ are the predicted parameters of $p(\hat{y}|\hat{z}) \sim \mathcal{N}(\mu,\sigma^2)$; and $z$, $\hat{z}$ are the hyper-latent representation and its quantized version, respectively.  As depicted in Fig.~\ref{fig:Hyper},
The $\hat{y}$ and $\hat{z}$ form the latent representation of the input $x$. The hyperprior thus bears the interpretation of factorizing the joint distribution $p(\hat{y},\hat{z})$ as $p(\hat{y}|\hat{z})p(\hat{z})$, where $p(\hat{y}|\hat{z})$ and $p(\hat{z})$ are assumed to be Gaussian and factorial, respectively. \emph{Hu et al.} \cite{coarse,benchmark} extend the idea to include more than one layer of hyperprior, leading to a factorization of $p(\hat{y},\hat{z}_1,\hat{z}_2,\ldots,\hat{z}_n)=p(\hat{y}|\hat{z}_1)p(\hat{z}_1|\hat{z}_2),\ldots,p(\hat{z}_n)$, where $\hat{z}_1,\hat{z}_2,\ldots,\hat{z}_n$ form a multi-layer hyperprior. In addition to the use of hyperprior, \emph{Minnen et al.}~\cite{ContextModel}, \emph{Lee et al.}~\cite{lee}, \emph{Chen et al.}~\cite{attn}, and \emph{Cheng et al.}~\cite{ContextGMM} incorporate an autoregressive prior by 2D~\cite{ContextModel,ContextGMM,lee} or 3D~\cite{attn} masked convolution~\cite{pixelrnn}, in order to utilize causal contextual information for better density estimation. In particular, \emph{Cheng et al.}~\cite{ContextGMM} model $p(\hat{y}|\hat{z})$ with a Gaussian mixture distribution instead of a Gaussian.

%the overall prior can be factorial into $p_{prior}=p(\hat{y}|\hat{z})p(\hat{z})$. Although the side information $(\hat{z})$ still uses the global statistics, the hyper-synthesis transform utilizes the side information to estimate an image dependant latent distribution to minimize the gap of the mismatched distributions.

%\subsection{Improvements on Autoencoder Transform}
\textbf{Enhanced Autoencoding Transform:} The capacity of the autoencoding transform determines its expressiveness. 
%Since the original autoencoder transformation in \cite{GoogleHyperPrior} only adopts simple convolutional and nonlinear GDN layers, it leaves some spaces for improvement. 
\emph{Chen et al.}~\cite{attn} add residual blocks to the autoencoder along with several non-local attention modules (NLAM). NLAM is shown to facilitate spatial bit allocation among coding areas of varied texture complexity. Unlike most of the VAE-based systems, which operate at image level, the block-based autoencoder in~\cite{recurrent} divides the input image into non-overlapping macroblocks, each of which contains multiple sub-blocks coded sequentially using recurrent-based analysis and synthesis transforms. It has the striking feature of allowing high degree of computational parallelism at macroblock level. In general, most autoencoders are not guaranteed to reconstruct the input perfectly even when no quantization is involved.  

%They firstly divided the image into non-overlapping blocks, and for each of the block, they further subdivide into four sub-blocks and encode them by using LSTM in the clockwise order starting from the top left to fully exploit and remove spatial redundancy. 

\vspace{-0.6em}
\subsection{Flow-based Image Compression}

\begin{figure}[t]
\begin{center}
\begin{subfigure}{0.2\textwidth}
    % include first image
    \centering
    %\vspace{-0.8em}
    \includegraphics[width=\linewidth]{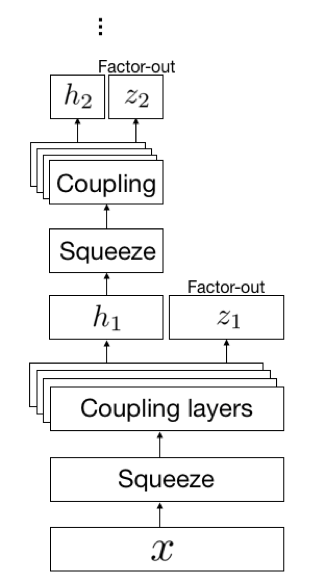} 
    \caption{Normalizing flows}
    %\vspace{-1.2em}
    \label{fig:NF}
\end{subfigure}
\begin{subfigure}{0.2\textwidth}
    % include first image
    \centering
    %\vspace{-0.8em}
    \includegraphics[width=\linewidth]{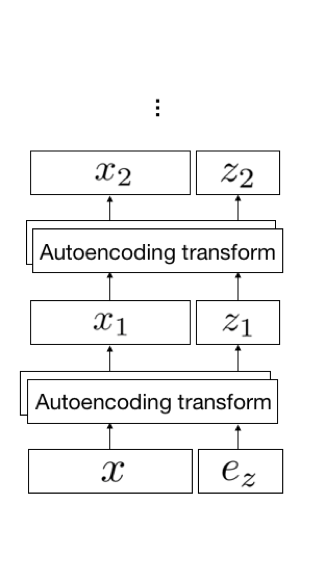}  
    \caption{ANFIC}
    %\vspace{-1.2em}
    \label{fig:oursANF}
\end{subfigure}

\caption{Flow-based image compression with (a) normalizing flows~\cite{NFcodec} and (b) augmented normalizing flows (ours).}
\label{fig:models}
\end{center}
\vspace{-2.0em}
\end{figure}

Recently, flow-based models~\cite{NF,nvp} emerge as an attractive alternative to VAE~\cite{vae} or other autoencoders. They are characterized by the bijective mapping between the input and its latent representation, ensuring that the input can be perfectly reconstructed from its latent in the absence of quantization.  
%Different from the commonly used autoencoder, another group of work changes the autoencoder to another type of invertible network []. 
\emph{Ma et al.}~\cite{iwave} make an interesting attempt to introduce lifting-based coupling layers, which are a specialized implementation of additive coupling layers~\cite{NF,nvp} often used to construct a flow model, as the analysis and synthesis backbone. In particular, they split an input image, first row-wise and then column-wise, into latent subbands, the resulting decomposition being similar to 2D wavelet transform.
%The resulting sub-bands are not guaranteed to have the same properties as the ordinary 2-D wavelet.    
%Next, they predict and unearth the correlation between the two parts to extract the high-frequency information. Finally, they exert the high-frequency information to update the low-frequency information. The whole process can be done on vertical and horizontal and repeated on the low low-frequency sub-band. Since wavelet-like transformation is volume-preserving, it does not incur information loss, and also it is invertible. 
\emph{Helminger et al.}~\cite{NFcodec} also use additive coupling layers but with the factor-out splitting to generate a multi-scale image representation as shown in Fig.~\ref{fig:NF}. Their work extends the notion of integer discrete flows for lossless compression~\cite{IDF} to lossy compression. In common, these works show the potential of flow-based models to offer a wide range of quality levels ranging from low-rate compression to nearly-lossless or even lossless compression. 
%introduced a norm propose an image codec based on moralizing flow.  wavelet-like transformation, they use coupling and factor out layers to simplify the representation by only further processing a part of the features.

Our work aims to leverage the developments of VAE-based schemes in a flow-based framework to enjoy the benefits of both (see Fig.~\ref{fig:oursANF}). For this purpose, we resort to augmented normalizing flows~\cite{ANF}, the basics of which are presented next.  
%different from AVE-based autoencoder transformation, we apply ANF () which is similar to normalizing flow but with the augmented space that will be explained in detail next.

\subsection{Augmented Normalizing Flows (ANF)}
\label{subsec:anf}

\begin{figure}[t]
\begin{center}
\begin{subfigure}{0.12\textwidth}
    % include first image
    \centering
    %\vspace{-0.8em}
    \includegraphics[width=\linewidth]{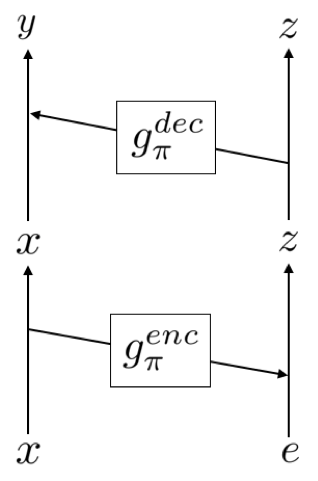} 
    \caption{}
    %\vspace{-1.2em}
    \label{fig:vae}
\end{subfigure}
\begin{subfigure}{0.22\textwidth}
    % include first image
    \centering
    %\vspace{-0.8em}
    \includegraphics[width=\linewidth]{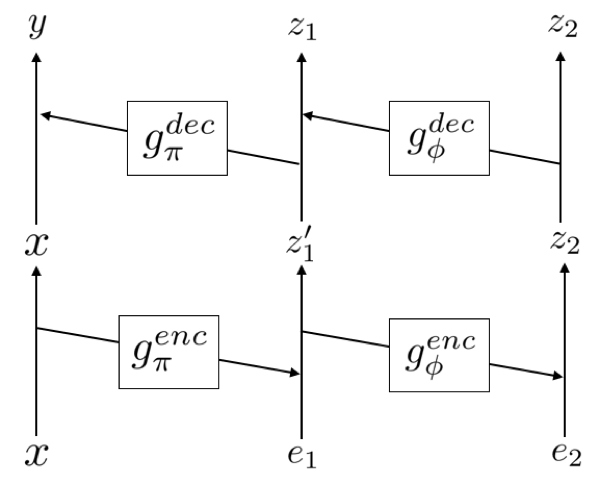}  
    \caption{}
    %\vspace{-1.2em}
    \label{fig:hanf}
\end{subfigure}

\caption{The architectures of ANF: (a) one-step ANF, composed of the encoding $g_\pi^{enc}$ and the decoding $g_\pi^{dec}$ transforms, and (b) one-step hierarchical ANF.}
\label{fig:autoencoding}
\end{center}
\vspace{-2.0em}
\end{figure}

The ANF model~\cite{ANF} is an invertible latent variable model. It is composed of multiple \textit{autoencoding} transforms, each of which comprises a pair of the encoding and decoding transforms as depicted in Fig.~\ref{fig:vae}. Consider the example of ANF with one autoencoding transform (i.e. one-step ANF). It converts the input $x$ coupled with an independent noise $e$ into their latent representation $(y,z)$ with one pair of encoding and decoding transforms:
\begin{align}
    & g^{enc}_\pi(x,e)=(x, s^{enc}_\pi(x) \odot e + m^{enc}_\pi(x))=(x,z) \label{eq:ANF_1} \\
    & g^{dec}_\pi(x,z)=((x-\mu^{dec}_\pi(z))/\sigma^{dec}_\pi(z), z)=(y,z) \label{eq:ANF_2}
%\label{eq:ANF_2}
\end{align}
where $s_{\pi}^{enc}$, $m_{\pi}^{enc}$, $\mu_{\pi}^{enc}$, and $\sigma_{\pi}^{enc}$ are element-wise affine transformation parameters. These learnable parameters are driven by the encoding and decoding neural networks, the weights of which are referred collectively to as $\pi$. Compared with ordinary flow models, ANF augments the input with an independent noise. It is shown in~\cite{ANF} that the  augmented  input  space  allows a smoother transformation to the required latent space.

%Compared with ordinary flow models, ANF operates on an input space augmented with noises. The augmented noises enable extra degree of freedom to transform the input space more freely into the required latent space

%Huang et al.~\cite{ANF} also claim and demonstrate that it can be hardly to perfectly transformed the data distribution into the prior distribution by using a 1-step ANF (aka VAE). However, it can be improved by stacking multiple-step of ANF, even though the weights are shared for all steps, the latent distribution could become closer to the prior distribution if we stack more ANF steps.

\textbf{Multi-step ANF and Hierarchical ANF:} From Fig.~\ref{fig:vae} and according to Eqs.~\eqref{eq:ANF_1} and~\eqref{eq:ANF_2}, the encoding $g^{enc}_\pi$ or decoding $g^{dec}_\pi$ transform implements an invertible affine coupling layer. Stacking pairs of these coupling layers leads to an invertible network, termed multi-step ANF, with much improved capacity than one-step ANF. Another way to increase the model capacity is to augment more noise inputs as hierarchical ANF (see Fig.~\ref{fig:hanf}). Particularly, these two approaches can be combined in a flexible way for even higher model capacity. 

%\textcolor{red}{To enhance the expressiveness of the prior distribution for complex input images, the latent variables are partitioned into disjoint groups, $z = \{z_1, z_1, ... , z_L\}$, and the joint distribution is presented by $p(x, z_1, z_1, ... , z_L) = p(x|z_1, z_1, ... , z_L) \prod_{l}^{L}p(z_l|z_{l+1}...z_L)$.}
 %Huang et al.~\cite{ANF} also claim and demonstrate that it can be hardly to perfectly transformed the data distribution into the prior distribution by using a 1-step ANF (aka VAE). However, it can be improved by stacking multiple-step of ANF, even though the weights are shared for all steps, the latent distribution could become closer to the prior distribution if we stack more ANF steps.
\textbf{Training ANF:} Like the ordinary flow models, ANF can be trained by maximizing the \textit{augmented joint likelihood}, i.e. $\arg\max_{\pi} p_\pi(x,e)$:
\begin{equation}
%\begin{split}
    p_{\pi}(x,e) = p(G_{\pi}(x,e))\left|det\frac{\partial G_{\pi }(x,e)}{\partial (x,e)}\right|,
\label{eq:ANF_3}
%\end{split}
\end{equation}
where $G_{\pi} = g^{dec}_{\pi_{N}} \circ g^{enc}_{\pi_{N}} \circ \ldots \circ g^{dec}_{\pi_1} \circ g^{enc}_{\pi_1}$ is the alternate composition of the encoding and decoding transforms with $\pi = \{\pi_{1},\cdot \cdot \cdot,\pi_{N} \}$ and $p(G_{\pi}(x,e))$ represents the specified or learned prior distribution over the latents $(y,z)$. It is shown in~\cite{ANF} that maximizing the augmented joint likelihood $p_\pi(x,e)$ in ANF amounts to maximizing a lower bound on the marginal likelihood $p_\pi(x)$, with the gap attributed to the model's incapability of modeling $e$ independently of $x$. 

\textbf{VAE as One-step ANF:} Notably, VAE can be viewed as one-step ANF by (1) letting $e \sim~ \mathcal{N}(0,I)$ be a Gaussian noise, (2) transforming $e$ into $z$ via re-parameterizing the VAE's encoding distribution $q^{enc}_\pi(z|x)$ of the form $\mathcal{N}(m^{enc}_\pi(x),(s^{enc}_\pi(x))^2)$, and (3) normalizing $x$ as $y=(x-\mu^{dec}_\pi(z))/\sigma^{dec}_\pi(z)$ via the VAE's decoding distribution $p^{dec}_\pi(x|z)=\mathcal{N}(\mu^{dec}_\pi(z),(\sigma^{dec}_\pi(z))^2)$. The resulting $y$ then follows $\mathcal{N}(0,I)$ and so does the aggregated distribution of $z$ from various inputs $x$. Maximizing Eq.~\eqref{eq:ANF_3} for such an one-step ANF is shown in~\cite{ANF} to be identical to maximizing the ELBO of VAE~\cite{vae}. 
\section{Proposed Method}
\label{sec:method}

Inspired by the fact that most learned image compression is VAE-based and that VAE is equivalent to one-step ANF, we propose an ANF-based image compression framework, termed ANFIC. We first outline the ANFIC framework in Section \ref{subsec:ANFIC}, with a focus on how to extend VAE-based image compression with hyperprior by multi-step and hierarchical ANF. This is followed by discussions on the entropy coding of the latent representation (Section \ref{sec:latent}), the modeling of the prior distribution in ANFIC (Section \ref{subsec:ANFIC}), and the training objective (Section \ref{Sec:TrainingObj}). 
%The last part of this section \textcolor{red}{(Section \ref{sec:Variable})} addresses how to achieve multi-rate compression with a single model.

To the best of our knowledge, ANFIC is the first work that combines VAE and flow models in a unified framework. It distinguishes from flow-based compression in that it operates on augmented input space (see Fig.~\ref{fig:oursANF}), being able to leverage the representation power of any existing VAE-based image compression. Moreover, ANFIC is more general than the VAE-based scheme by allowing it to be stacked and/or extended hierarchically (see Fig.~\ref{fig:autoencoding}).

\begin{figure}[t!]
\begin{center}
\begin{subfigure}{0.23\textwidth}
    % include first image
    \centering
    %\vspace{-0.8em}
    \includegraphics[width=\linewidth]{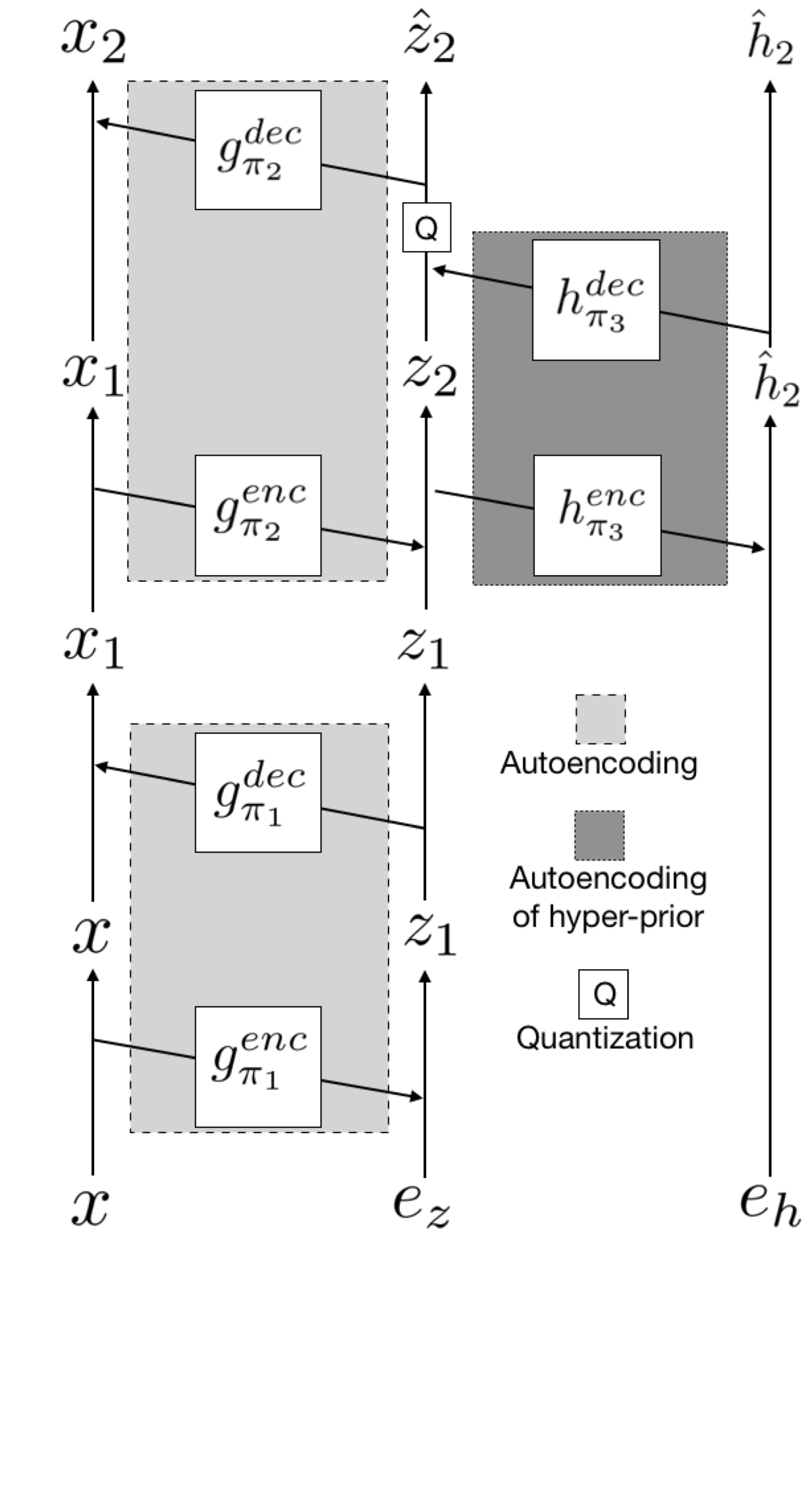} 
    \caption{}
    %\vspace{-1.2em}
    \label{fig:overall}
\end{subfigure}
\begin{subfigure}{0.24\textwidth}
    % include first image
    \centering
    %\vspace{-0.8em}
    \includegraphics[width=\linewidth]{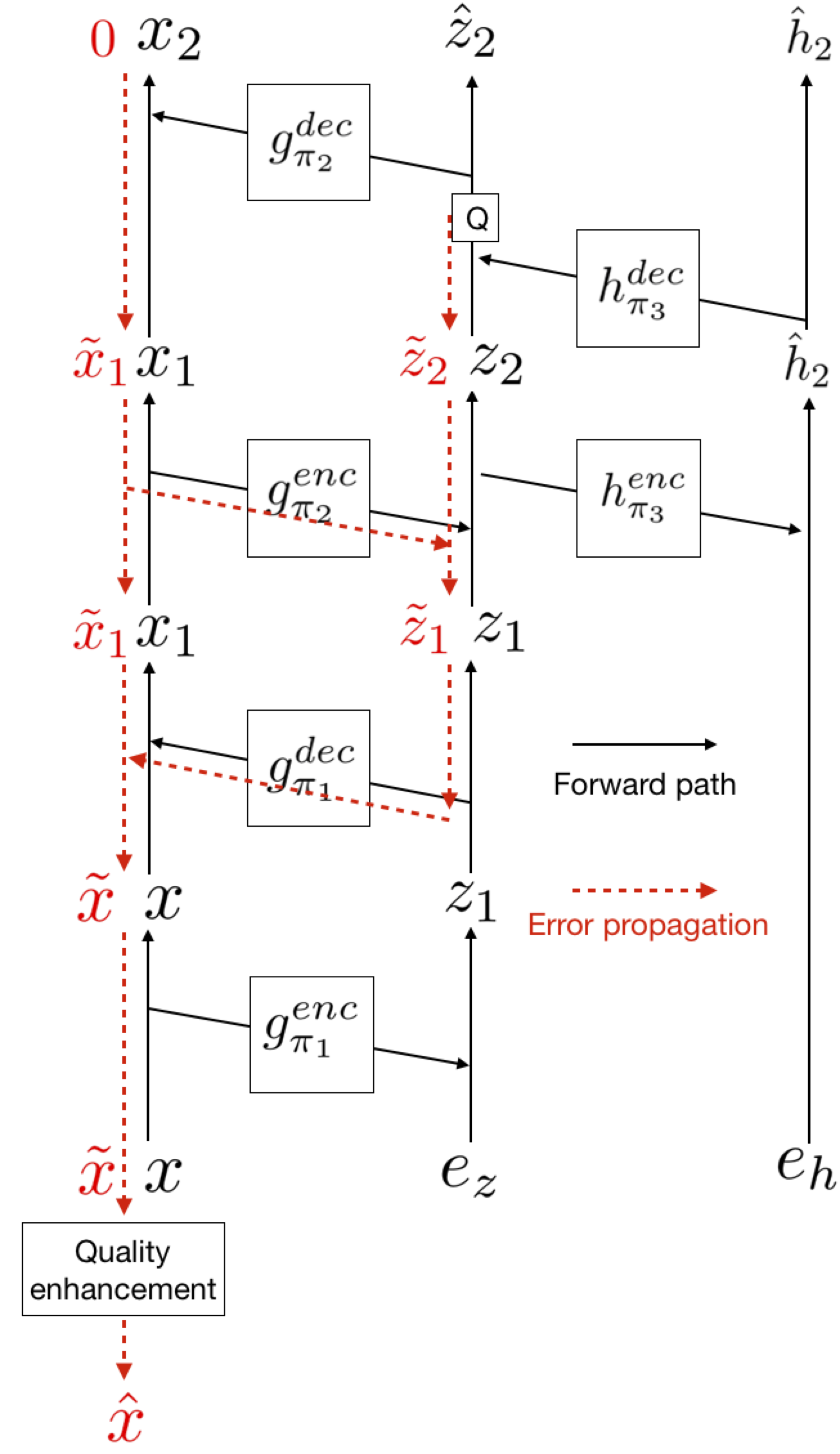}
    \caption{}
    %\vspace{-1.2em}
    \label{fig:error}
\end{subfigure}

\caption{(a) The overall architecture of our proposed ANF-based image compression (ANFIC). (b) Error propagation due to the quantization of the image latents $x_2, z_2$. To alleviate propagation errors, we place a quality enhancement (QE) network at the end of the reverse path (the red dotted line).}
\label{fig:overallerror}
\end{center}
\vspace{-2.0em}
\end{figure}

\begin{comment}
\begin{figure}[t!] %!t
\centering
\includegraphics[width=0.5\linewidth]{FIG/overall.png}
\caption{The overall architecture of our proposed ANF-based image compression (ANFIC). }
\label{fig:overall}
\vspace{-0.6em}
\end{figure}
\end{comment}

\subsection{ANFIC Framework}
\label{subsec:ANFIC}
Fig.~\ref{fig:overall} describes the framework of ANFIC. From bottom to top, it stacks two autoencoding transforms (i.e.~two-step ANF), with the top one extended further to the right to form a hierarchical ANF~\cite{ANF} that implements the hyperprior. More autoencoding transforms can be added straightforwardly to create a multi-step ANF. In particular, the $g^{enc}_{\pi}$ and $g^{dec}_{\pi}$ in the autoencoding transform follow Eqs.~\eqref{eq:ANF_1} and~\eqref{eq:ANF_2}, except that we make them purely additive by removing $s^{enc}_\pi(x)$ and $\sigma^{dec}_\pi(z)$ for better convergence as with some other flow-based schemes~\cite{NFcodec, iwave}. 

The autoencoding transform of the hyperprior, which assumes each sample in the latent representation $z_2$ is a Gaussian, is defined as
\begin{align}
    & h^{enc}_{\pi_3}(z_2,e_h)=(z_2, e_h + m^{enc}_{\pi_3}(z_2))=(z_2,\hat{h}_2), \label{eq:ANF_hp_p1}\\
    & h^{dec}_{\pi_3}(z_2,\hat{h}_2)=(\lfloor z_2 - \mu_{\pi_3}^{dec}(\hat{h}_2)  \rceil, \hat{h}_2)=(\hat{z}_2, \hat{h}_2),
\label{eq:ANF_hp}
\end{align}

%\textcolor{blue}{where $Quant$ (depicted as Q in Fig. \ref{fig:overallerror}) simulates the additive quantization noise during training and denotes the nearest-integer rounding at inference time for quantizing the residual between $z_2$ and the predicted mean $\mu_{\pi_3}^{dec}(\hat{h}_2)$ of the Gaussian distibution from the hyperprior $\hat{h}_2$.}
where $\lfloor \cdot \rceil$ (depicted as Q in Fig. \ref{fig:overall}) denotes the nearest-integer rounding for quantizing the residual between $z_2$ and the predicted mean $\mu_{\pi_3}^{dec}(\hat{h}_2)$ of the Gaussian distibution from the hyperprior $\hat{h}_2$. 
This part implements the autoregressive hyperprior in~\cite{ContextModel}, with $z_2$ denoting the image latents whose distributions are signaled as the side information $\hat{h}_2$.

The encoding of ANFIC proceeds by passing the augmented input $(x, e_z, e_h)$ through the autoencoding and hyperprior transforms, i.e. $G_{\pi} = g^{dec}_{\pi_2}  \circ  h^{dec}_{\pi_3} \circ h^{enc}_{\pi_3} \circ g^{enc}_{\pi_2} \circ g^{dec}_{\pi_1} \circ g^{enc}_{\pi_1}$, to obtain the latent representation $(x_2, \hat{z}_2, \hat{h}_2)$. In particular, $x$ represents the input image, $e_z = 0$ denotes the augmented input%\sim \mathcal{N}(0,I)$ denotes the augmented Gaussian noise
, and $e_h \sim \mathcal{U}(-0.5,0.5)$, another augmented input, simulates the additive quantization noise of the hyperprior during training. To achieve lossy compression, we want $\hat{z}_2$ and $\hat{h}_2$ to capture most of the information about the input $x$ and regularize $x_2$ during training to approximate noughts. As such, only $\hat{z}_2$ and $\hat{h}_2$ are entropy coded into bitstreams. Note that due to the volume-preserving property of ANF (or any flow model), $x_2$ has the same dimensionality as the input $x$ while that of $\hat{z}_2$ and $\hat{h}_2$ is usually much smaller depending on the design choice. This flexibility allows us to incorporate any existing VAE-based compression scheme as one specific realization of the autoencoding transform in ANFIC. For example, the encoder of any VAE-based compression can be used to implement $m^{enc}_\pi(x)$ for the encoding transform in Eq.~\eqref{eq:ANF_1}; likewise, its decoder can  realize $\mu^{dec}_\pi(x)$ for the decoding transform in Eq.~\eqref{eq:ANF_2}. Note that we have assumed the use of additive coupling layers.

To decode the input $x$, we apply the inverse mapping function $G^{-1}_{\pi}$ to the quantized latents $(0, \hat{z}_2, \hat{h}_2)$, where $x_2$ is set to noughts. In ANFIC, there are two sources of distortion that cause the reconstruction to be lossy: the quantization error of $z_2$ and the error of setting $x_2$ to noughts during the inverse operation. Essentially, ANFIC is an ANF model, which is bijective and invertible. The errors between the encoding latents $(x_2, z_2)$ and their quantized version $(0, \hat{z}_2)$ will introduce distortion to the reconstructed image, as shown in Fig.~\ref{fig:error}. 

To mitigate the effect of quantization errors on the decoded image quality, we incorporate a quality enhancement (QE) network at the end of the reverse path, as illustrated in Fig.~\ref{fig:error}. This enhancement network is an integral part of ANFIC, which is constrained by the fact that the analysis and the synthesis transforms must share the same autoencoding transforms (i.e. invertible coupling layers). This constraint makes it difficult to learn a synthesis transform that can effectively compensate for quantization errors while maintaining the invertibility. The same observation was made in~\cite{iwave}. In this paper, we adopt the same lightweight quality enhancement network as~\cite{iwave}.    
%Different from VAE-based methods that their decoders are flexible enough to have the ability to overcome the quantization error, the quantization error introduced by ANFIC after the forward transform can hardly be compensated via the inverse transformation. To this end, 

\begin{comment}
\begin{figure}[t!] %!t
\centering
\includegraphics[width=0.55\linewidth]{FIG/error.png}
\caption{Error propagation due to the quantization of the image latents $x_2, z_2$. To alleviate propagation errors, we place a quality enhancement network at the end of the reverse path (the red dotted line).}
\label{fig:error}
\vspace{-0.3em}
\end{figure}
\end{comment}

\subsection{Prior Distribution}
\label{sec:latent}
The prior distribution of ANFIC refers to the joint distribution $p(x_2, \hat{z}_2, \hat{h}_2)$ of the latents $(x_2, \hat{z}_2, \hat{h}_2)$, which like VAE-based schemes plays a crucial role in determining the rate needed to signal the image latents. Rather than manually specifying the prior distribution, we adopt a parametric approach to learn $p(x_2, \hat{z}_2, \hat{h}_2)$, for the sake of balancing between rate and distortion. As noted previously, our ANFIC has the latent $\hat{z}_2$ and the hyperprior $\hat{h}_2$ capture most of the information of the input $x$. We thus regularize the latent $x_2$ to follow a zero-mean Gaussian with a small variance $\sigma^2$ and to be independent of $\hat{z}_2,\hat{h}_2$. 
That is, $p(x_2, \hat{z}_2, \hat{h}_2)$ factorizes as: 
\begin{equation}
     p(x_2, \hat{z}_2, \hat{h}_2) = p(x_2) p(\hat{z}_2|\hat{h}_2) p(\hat{h}_2),
\label{eq:ANF_hp1}
%\vspace{-0.3cm}
\end{equation}
with  
\begin{equation}
     p(x_2) = \mathcal N(0,\sigma^2), 
\end{equation}
 and the remaining terms, $p(\hat{z}_2|\hat{h}_2)$ and $p(\hat{h}_2)$, learned from data by neural networks. 
 
Similar to VAE-based schemes \cite{GoogleHyperPrior}, we assume $p(\hat{h}_2)$ to be a non-parametric distribution and $p(\hat{z}_2|\hat{h}_2)$ to be a conditional Gaussian. Recall that $\hat{z}_2$ and $\hat{h}_2$ are the quantized version of the primary image latent $z_2$ and its hyperprior $m^{enc}_{\pi_3}(z_2)$ (see Eq.~\eqref{eq:ANF_hp_p1}), which is output by the encoding transform of $h_{\pi_3}^{enc}$ (see Fig.~\ref{fig:overall}). We follow the additive noise model for quantization during training. As a result, we have $\hat{h}_2=m^{enc}_{\pi_3}(z_2)+e_h,e_h \sim \mathcal{U}(-0.5,0.5)$ and $\hat{z}_2 = \lfloor z_2 - \mu_{\pi_3}^{dec}(\hat{h}_2) \rceil$
%$\hat{z}_2 = \lfloor z_2 - \mu_{\pi_3}^{dec}(\hat{h}_2) \rceil$ 
follow a distribution given by the convolution of $\mathcal{N}(0, (\sigma_{\pi3}^{dec}(\hat{h}_2))^2)$ and $\mathcal{U}(-0.5,0.5)$.  In symbols, $p(\hat{h}_2)$ and $p(\hat{z}_2|\hat{z}_2)$ have the forms of
\begin{equation}
\begin{split}
     %p(x_2) &= \mathcal N(0,\sigma^2) \\
     p(\hat{z}_2|\hat{h}_2) &= \mathcal{N}(0,(\sigma^{dec}_{\pi_3}(\hat{h}_2))^2) \ast \mathcal{U}(-0.5,0.5) \\
     p(\hat{h}_2) &= \mathcal P_{\hat{h}_2|\psi} \ast \mathcal{U}(-0.5,0.5)
\label{eq:ANF_hp2}
%\vspace{-0.1cm}
\end{split}
\end{equation}
where $\ast$ denotes convolution and $P_{\hat{h}_2|\psi}$ is a learned distribution parameterized by $\psi$. Note that unless otherwise specified, $p(x_2),p(\hat{z}_2|\hat{h}_2),p(\hat{h}_2)$ are all assumed to be factorial over the elements of $x_2,\hat{z}_2,\hat{h}_2$, respectively.

Algorithm 1 and 2 present the encoding and decoding procedures of ANFIC, respectively, where the $\tilde{x}_1$, $\tilde{z}_2$, and $\tilde{x}$ stand for the reconstructed version of $x_1$, $x_2$, and $x$, respectively (See Fig.~\ref{fig:error}).

\begin{algorithm}[t]
\begin{footnotesize}
\caption{The encoding procedure of ANFIC}
\label{alg:enc}
\begin{algorithmic}[1]
\STATE Input: The image $x$ and the augmented inputs $e_z,e_h$
\STATE Output: The bitstream of $\hat{z}_2$ and $\hat{h}_2$
\STATE $z_1 = m^{enc}_{\pi_1}(x) + e_z$
\STATE $x_1 = x - \mu^{dec}_{\pi_1}(z_1)$
\STATE $z_2 = z_1 + m^{enc}_{\pi_2}(x_1)$
\STATE $\hat{h}_2 = m^{enc}_{\pi_3}(z_2) + e_h$ (replaced with the nearest-integer rounding of $m^{enc}_{\pi_3}(z_2)$ at inference time)
\STATE Encode $\hat{h}_2$ using $p(\hat{h}_2)$ in Eq.~\eqref{eq:ANF_hp2} \STATE $\hat{z}_2 = \lfloor z_2 - \mu^{dec}_{\pi_3}(\hat{h}_2) \rceil$
\STATE Encode $\hat{z}_2$ using $p(\hat{z}_2|\hat{h}_2)$ in Eq.~\eqref{eq:ANF_hp2}
\STATE $x_2 = x_1 - \mu^{dec}_{\pi_2}(\hat{z}_2)$
\end{algorithmic}
\end{footnotesize}
\end{algorithm}

\begin{algorithm}[t]
\begin{footnotesize}
\caption{The decoding procedure of ANFIC}
\label{alg:dec}
\begin{algorithmic}[1]
\STATE Input: The bitstream of $\hat{z}_2$ and $\hat{h}_2$
\STATE Output: The reconstructed image $\hat{x}$
\STATE Set $x_2$ to 0
\STATE Decode $\hat{h}_2$ using $p(\hat{h}_2)$ in Eq.~\eqref{eq:ANF_hp2} 
\STATE Decode $\hat{z}_2$ using $p(\hat{z}_2|\hat{h}_2)$ in Eq.~\eqref{eq:ANF_hp2}
\STATE $\tilde{x}_1 = \mu^{dec}_{\pi_2}(\hat{z}_2)$
\STATE $\tilde{z}_2 = \hat{z}_2 + \mu^{dec}_{\pi_3}(\hat{h}_2)$
\STATE $\tilde{z}_1 = \tilde{z}_2 - m^{enc}_{\pi_2}(\tilde{x}_1)$
\STATE $\tilde{x} = \tilde{x}_1 + \mu^{dec}_{\pi_1}(\tilde{z}_1)$
\STATE $\hat{x} = QE(\tilde{x})$
\end{algorithmic}
\end{footnotesize}
\end{algorithm}

%Additionally, we can further extend the single Gaussian model into a Gaussian mixture model to enhance the capability of multi-model distribution estimation. 
\textbf{Gaussian Mixtures Extension:} ANFIC is flexible in accommodating more sophisticated modeling of $p(\hat{z}_2|\hat{h}_2)$, such as Gaussian mixture models. Unlike the single Gaussian model, the mixture model requires to estimate the mixing probabilities $w_{(k)},k=1,2,\ldots,K$ for $K$ components as well as the corresponding mean $\mu_{(k)}$ and variance $\sigma_{(k)}$. All these parameters are functions of the hyperprior $\hat{h}_2$. In the present case, the decoding transform $h^{dec}_{\pi_3}$ (see Eq.~\eqref{eq:ANF_hp}) is changed to $h^{dec}_{\pi_3}(z_2,\hat{h}_2)=(\lfloor z_2  \rceil, \hat{h}_2)=(\hat{z}_2, \hat{h}_2)$--namely, an identity transform followed by the quantization of $z_2$. This change is necessary because with the mixture model, the subtraction of a single predicted mean from $z_2$ is not feasible. In addition, $p(\hat{z}_2|\hat{h}_2)$ follows a distribution given by
\begin{equation}
\begin{split}
p(\hat{z}_2|\hat{h}_2) = & \left( \sum_{k=1}^K w_{(k)}(\hat{h}_2) \mathcal{N}(\mu^{dec}_{\pi_3(k)}(\hat{h}_2),(\sigma^{dec}_{\pi_3(k)}(\hat{h}_2))^2) \right)   \\ &  \ast \mathcal{U}(-0.5,0.5) %\\
\end{split}
\end{equation}

%Likewise, the choice for $p(\hat{h}_2)$ is flexible. 
%the convolution of $\sum_{k=1}^K w_{(k)} \mathcal{N}(\mu^{dec}_{\pi_3(k)}(\hat{h}_2),(\sigma^{dec}_{\pi_3(k)}(\hat{h}_2))^2)$ and $\mathcal{U}(-0.5,0.5)$ and $p(\hat{z}_2|\hat{h}_2)$ in Eq.~\eqref{eq:ANF_hp2} changes to:

\vspace{-0.8em}
\subsection{Training Objective}
\label{Sec:TrainingObj}
Training ANFIC can be achieved by minimizing the negative augmented log-likelihood, i.e. $\arg\min_{\pi,\psi} -\log p_{\pi,\psi}(x,e_z,e_h)$. This leads to the following loss function:
\begin{equation}
\begin{split}
    \mathcal{L}(x, e_z, e_h;\pi,\psi)  = &
    -\log p(\hat{h}_2) -\log p(\hat{z}_2|\hat{h}_2) + \lambda_1 \|x_2-0\|^2 \\
    & -log\left|det\frac{\partial G_{\pi }(x,e_z,e_h)}{\partial (x,e_z,e_h)}\right|,  
\label{eq:weighted_prediction}
\end{split}
\vspace{-0.3cm}
\end{equation} 
where the Jacobian log-determinant generally prevents the collapse of the latent space. In our implementation, we replace it with a reconstruction loss $\lambda_2 d(x,\hat{x})$, with the distortion metric $d(\cdot,\cdot)$ being the mean-squared error (MSE) or multi-scale structure similarity index (MS-SSIM):
\begin{equation}
\begin{split}
    & \mathcal{L}(x, e_z, e_h;\pi,\psi)  \\ 
    & = 
    \underbrace{-\log p(\hat{h}_2) -\log p(\hat{z}_2|\hat{h}_2)}_R + \lambda_1 \|x_2-0\|^2 
     + \underbrace{\lambda_2 d(x,\hat{x})}_D,  
\label{eq:training_objective}
\end{split}
\vspace{-0.3cm}
\end{equation} 
where $\pi,\psi$ refer to the parameters of all the networks, including the quality enhancement network. Unlike the traditional weighted sum of rate $R$ and distortion $D$, our training objective has the additional requirement that $x_2$ should approximate noughts. This drives $\hat{z}_2,\hat{h}_2$ to encode most of the information about the input $x$, provided that the reconstructed image $\hat{x}$ approximate $x$ closely. In passing, we note that the reconstruction loss also prevents the latent space from collapsing. Apparently, it would be difficult to recover the input $x$ if different $x$'s are all mapped to the same point in the latent space. 

\section{Experimental Results}
\label{sec:experiment}

This section evaluates the performance of ANFIC both objectively and subjectively. We first present the network architectures, training details, evaluation methodologies, and the baseline methods in Section~\ref{subsec:Settings}. Next, we compare the rate-distorton performance of ANFIC with several state-of-the-art methods on commonly used datasets in Section~ \ref{subsec:rd}. Lastly, we evaluate the subjective quality of the reconstructed images in Section~\ref{subsec:subjective}.

\begin{figure}[t!] %!t
\centering
\includegraphics[width=0.85\linewidth]{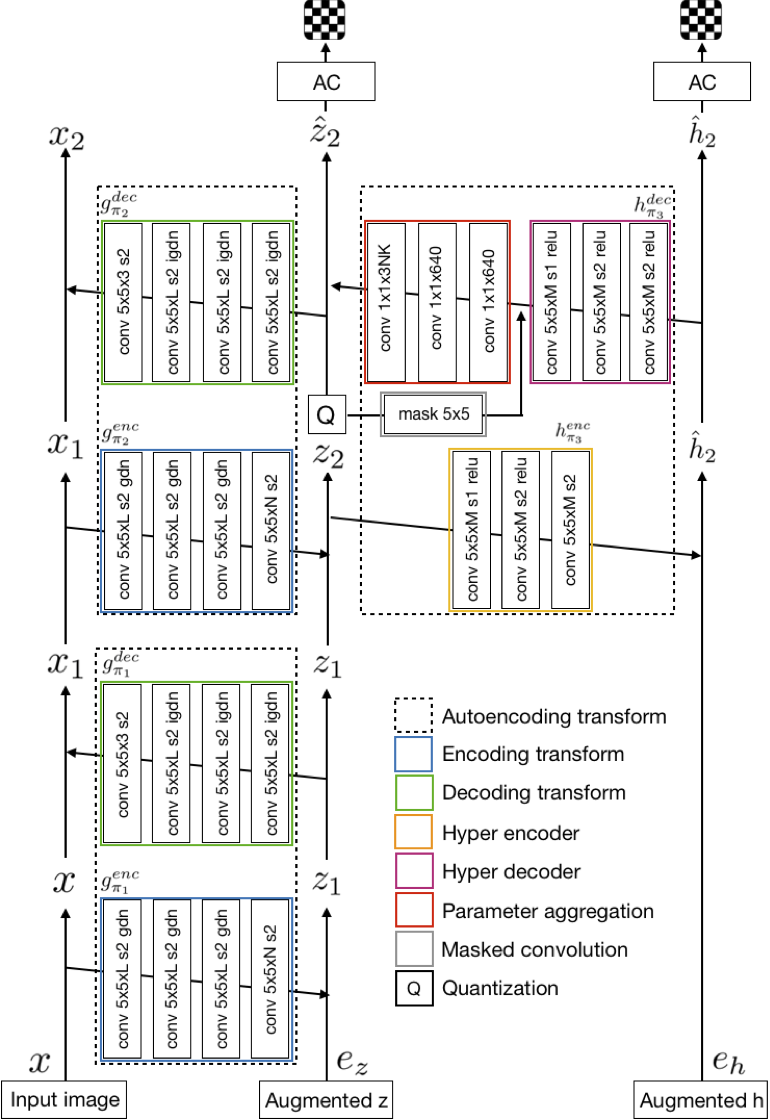}
\caption{The network architecture of our proposed ANFIC $(L=128, N=320, M=192, K=3)$. We adopt the autoregressive and Gaussian mixture model for entropy coding. AC and mask denote arithmetic coding and masked convolution, respectively.}
\label{fig:architecture}
%\vspace{-2.0em}
\end{figure}

\subsection{Settings and Implementation Details}
\label{subsec:Settings}

\emph{\textbf{Network Architectures:}}
Our autoencoding transforms for feature extraction (the left branch in Fig.~\ref{fig:architecture}) and hyperprior (the right branch in Fig.~\ref{fig:architecture}) share similar architectures to the VAE-based scheme in~\cite{ContextModel}. In addition, we use the same lightweight de-quantization network in~\cite{iwave} as the quality enhancement network. All the autoencoding transforms in our model have separate network weights. To keep the overall model size comparable to that of~\cite{ContextModel}, we reduce the number of channels in every convolutional layer to 128. We adopt the autoregressive and Gaussian mixture model (Section~\ref{sec:latent}) for entropy coding in all the experiments, with the number $K$ of mixture components set empirically to 3, which is found to be most effective in~\cite{ContextGMM}.

\emph{\textbf{Training:}} 
For training, we use \emph{vimeo-90k} dataset from \cite{Vimeo90k}. It contains 91,701 training videos, each having 7 frames. In a training iteration, we randomly choose one frame from each video and crop it to $256 \times 256$. We adopt the Adam \cite{Adam} optimizer with a batch size of 32. The learning rate is fixed at $1e^{-4}$ during the first 3M iterations, and then we decay to $1e^{-5}$ for fine-tuning. The two hyper-parameters (see Eq.~\eqref{eq:training_objective}) are chosen to have $\lambda_1 = 0.01* \lambda_2$, where $\lambda_2$ is one of the values from $\{0.1, 0.05, 0.02, 0.01, 0.005, 0.002\}$ for MSE optimization and from $\{200, 100, 40, 20, 10, 4\}$ for optimizing MS-SSIM. In particular, we first train our model for the highest rate point. It is then fine tuned with few epochs to obtain the models for lower rate points.

\emph{\textbf{Evaluation:}} We evaluate our model on commonly used datasets, \emph{Kodak} \cite{Kodak} and \emph{Tecnick} \cite{Tecnick}, which include 24 uncompressed images of size $768 \times 512$ and 40 images of size $1200 \times 1200$, respectively. Additionally, we test our model on the CLIC validation datasets \cite{CLIC2020}. It contains two subdivided datasets: professional and mobile. The former has 41 higher resolution images and the latter 61 images. To evaluate the rate-distortion performance, we report rates in bits per pixel (bpp) and quality in PSNR-RGB and MS-SSIM. Moreover, we use BPG as an anchor in reporting the BD-rates. Note that rate inflation as compared to BPG is reflected by positive BD-rates while rate saving is shown as negative BD-rates.

\emph{\textbf{Baselines:}} For comparison, the baseline methods include VTM-444, BPG-444, ICLR'18~\cite{GoogleHyperPrior}, NIPS'18~\cite{ContextModel}, ICLR'19~\cite{lee}, TPAMI'20~\cite{iwave}, CVPR'20~\cite{ContextGMM}, TPAMI'21~\cite{benchmark}, and TIP'21~\cite{attn}. It is worth noting that TPAMI'20~\cite{iwave} is a flow-based model, while the other learned codecs are VAE-based.
\vspace{-0.8em}

\subsection{Rate-Distortion Performance}
\label{subsec:rd}

\begin{figure*}[t!]
\begin{center}
\begin{subfigure}{0.46\textwidth}
    % include first image
    \centering
    %\vspace{-0.8em}
    \includegraphics[width=\linewidth]{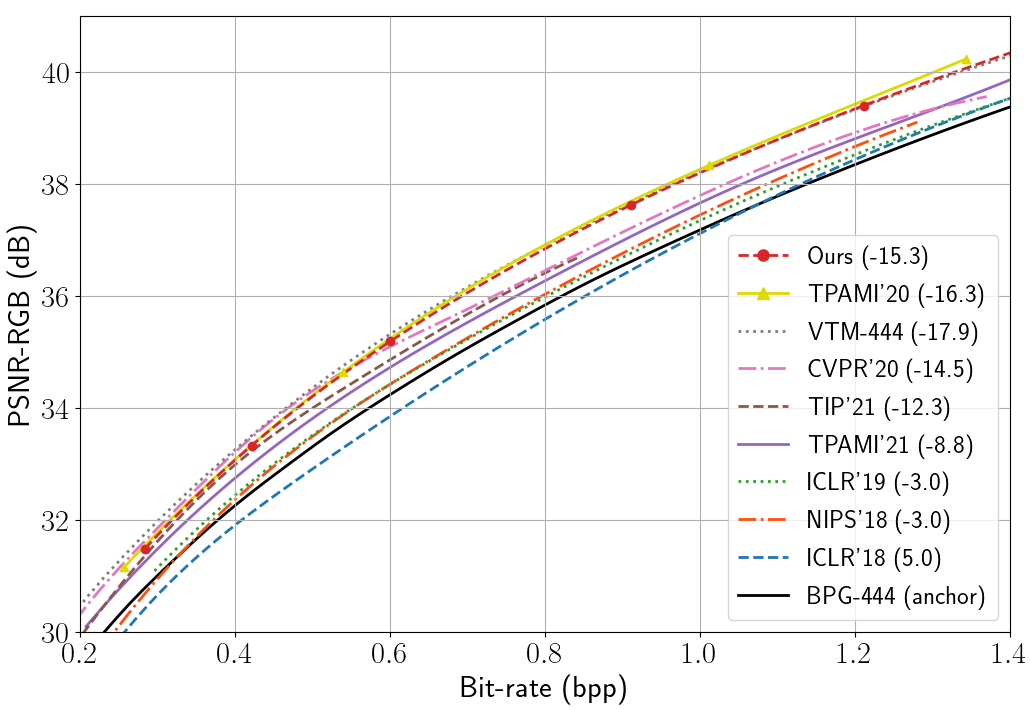} 
    \caption{Kodak, PSNR}
    %\vspace{-1.2em}
    \label{fig:KodakPSNR}
\end{subfigure}
\begin{subfigure}{0.46\textwidth}
    % include first image
    \centering
    %\vspace{-0.8em}
    \includegraphics[width=\linewidth]{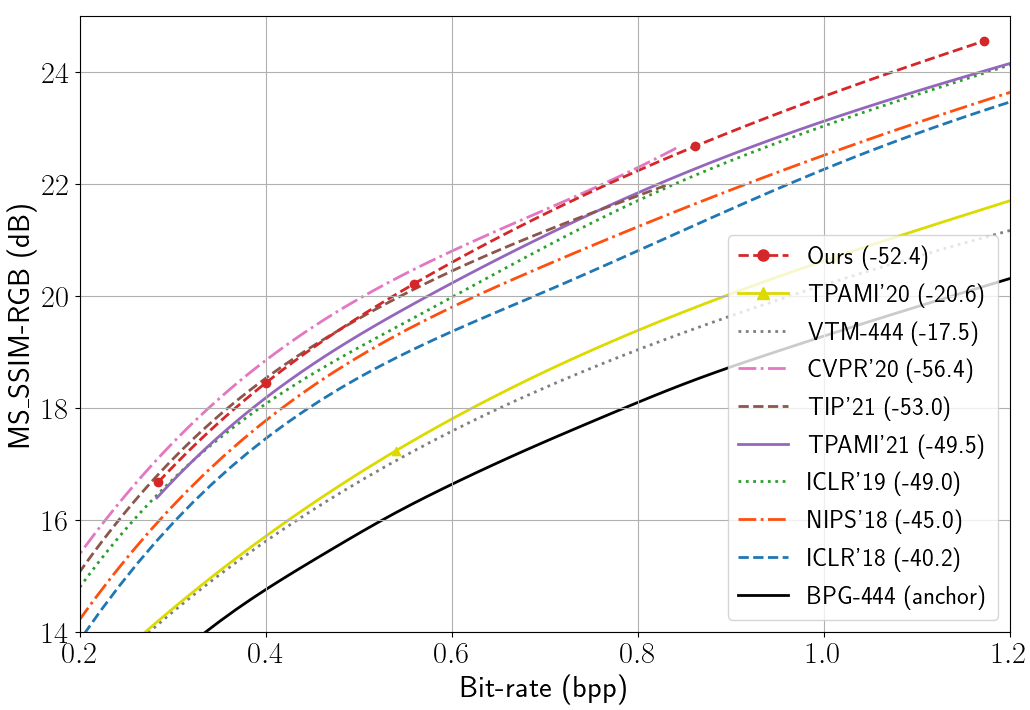}
    \caption{Kodak, MS-SSIM}
    %\vspace{-1.2em}
    \label{fig:KodakSSIM}
\end{subfigure}
\begin{subfigure}{0.46\textwidth}
    % include first image
    \centering
    %\vspace{-0.8em}
    \includegraphics[width=\linewidth]{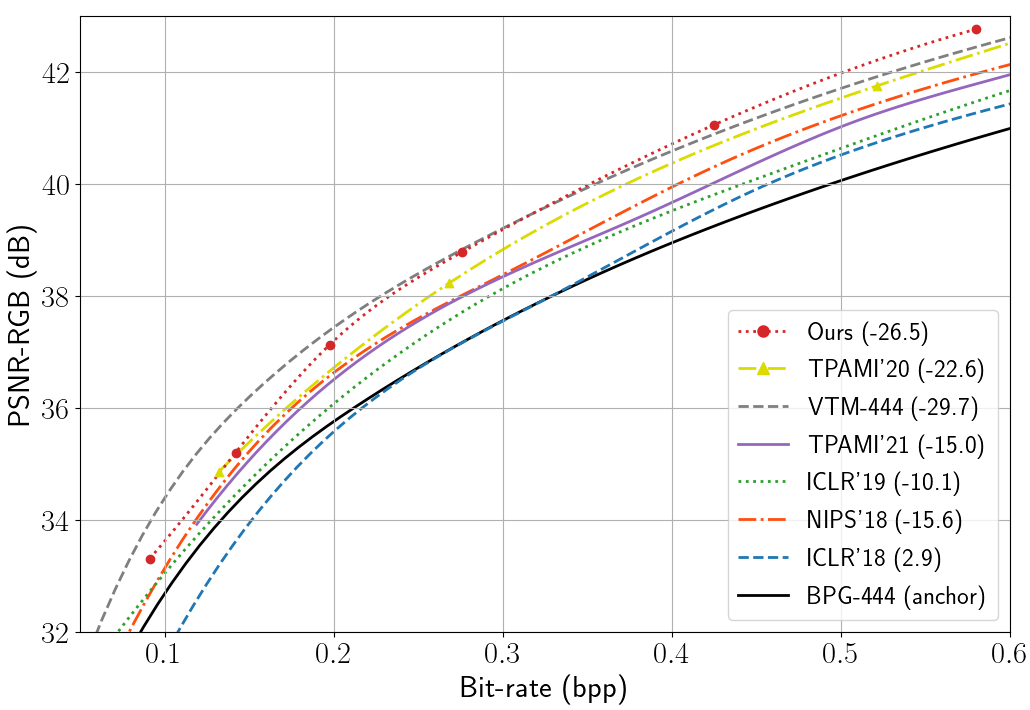}  
    \caption{Tecnick, PSNR}
    %\vspace{-1.2em}
    \label{fig:TecnickPSNR}
\end{subfigure}
\begin{subfigure}{0.46\textwidth}
    % include first image
    \centering
    %\vspace{-0.8em}
    \includegraphics[width=\linewidth]{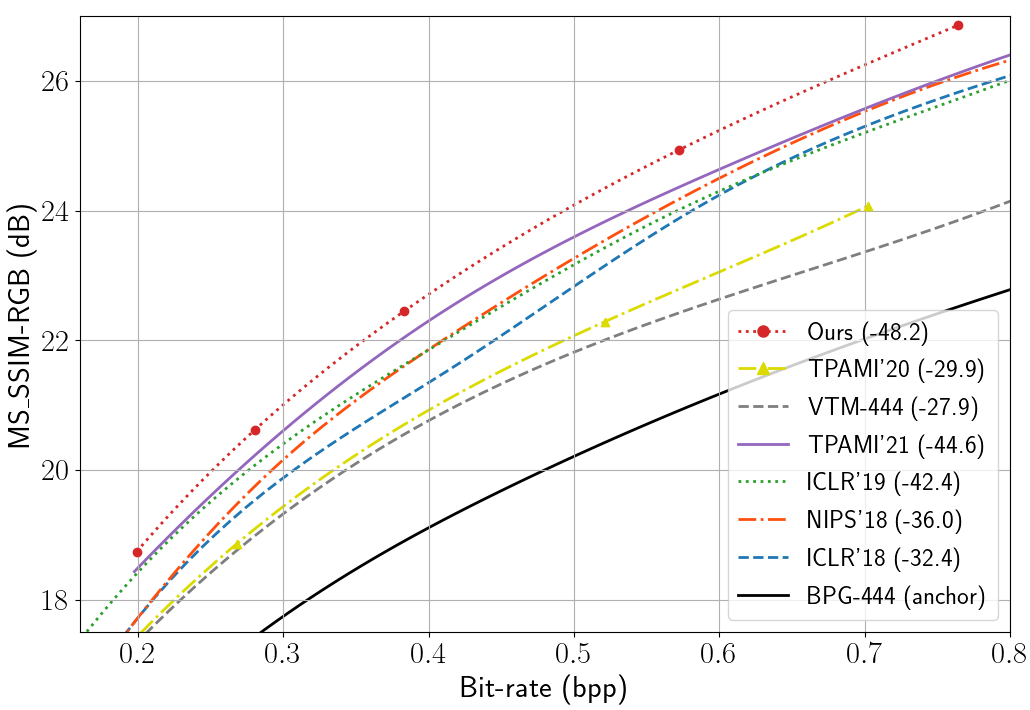} 
    \caption{Tecnick, MS-SSIM}
    %\vspace{-1.2em}
    \label{fig:TecnickSSIM}
\end{subfigure}
\begin{subfigure}{0.46\textwidth}
    % include first image
    \centering
    %\vspace{-0.8em}
    \includegraphics[width=\linewidth]{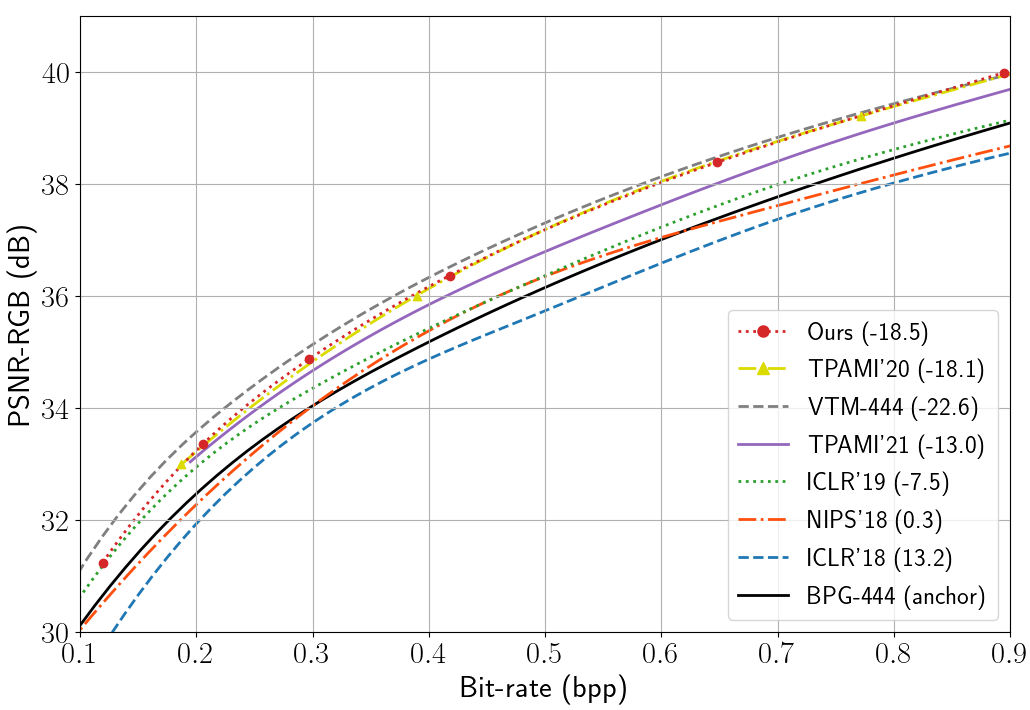}
    \caption{CLIC, PSNR}
    %\vspace{-1.2em}
    \label{fig:CLICPSNR}
\end{subfigure}
\begin{subfigure}{0.46\textwidth}
    % include first image
    \centering
    %\vspace{-0.8em}
    \includegraphics[width=\linewidth]{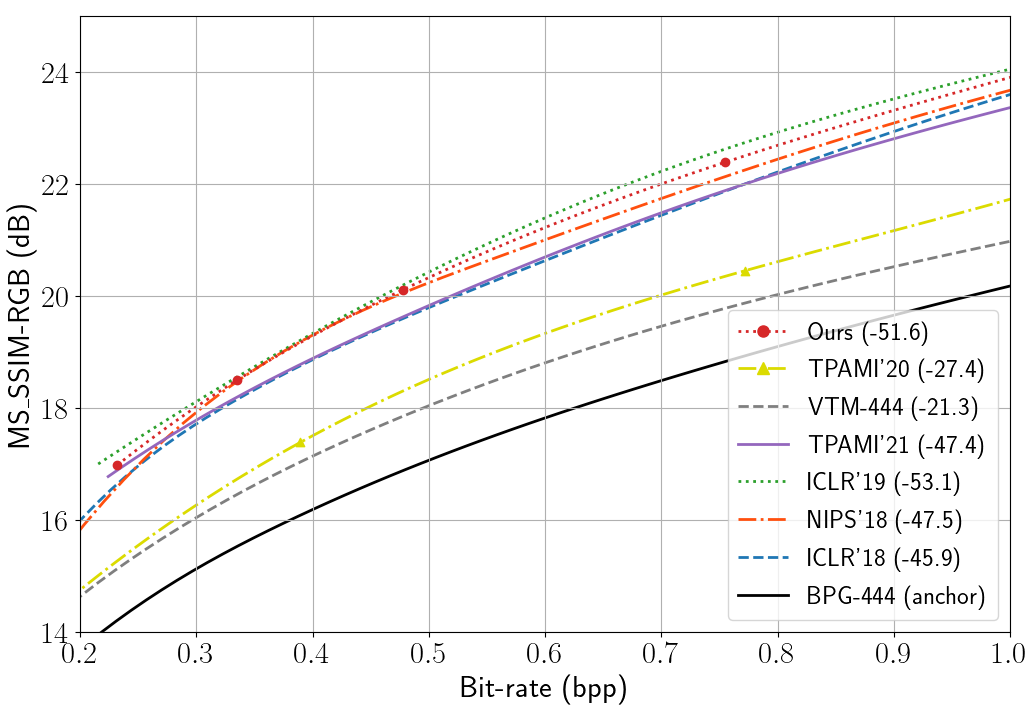}  
    \caption{CLIC, MS-SSIM}
    %\vspace{-1.2em}
    \label{fig:CLICSSIM}
\end{subfigure}
%\begin{subfigure}{0.45\textwidth}
%    % include first image
%    \centering
%    %\vspace{-0.8em}
%    \includegraphics[width=\linewidth]{FIG/R_D_CLIC_pro_PSNR.png}
%    \caption{CLIC-professional, PSNR}
%    %\vspace{-1.2em}
%    \label{fig:CLICproPSNR}
%\end{subfigure}
%\begin{subfigure}{0.45\textwidth}
%    % include first image
%    \centering
%    %\vspace{-0.8em}
%    \includegraphics[width=\linewidth]{FIG/R_D_CLIC_pro_SSIM.png}
%    \caption{CLIC-professional, MS-SSIM}
%    %\vspace{-1.2em}
%    \label{fig:CLICproSSIM}
%\end{subfigure}

%\caption{Rate-distortion performance evaluation on (a) Kodak (PSNR), (b) Kodak (MS-SSIM), (c) Techinck (PSNR), (d) Techinck (MS-SSIM), (e) CLIC (PSNR), and (f) CLIC (MS-SSIM). The numbers in the parentheses are the BD-rates with BPG-444 as anchor.}
\caption{Rate-distortion performance evaluation on Kodak, Tecnick, and CLIC datasets for both PSNR and MS-SSIM. The numbers in the parentheses are the BD-rates with BPG-444 as anchor.}
\label{fig:RD}
\end{center}
\vspace{-2.0em}
\end{figure*}

Fig.~\ref{fig:RD} compares the rate-distortion performance of the competing methods on Kodak, Tecnick, and CLIC (professional and mobile combined) datasets, with the BD-rate numbers summarized in Table~\ref{tab:exp_BD_PSNR}. Following some prior works, the BD-rate figures for CLIC professional dataset are reported separately in Table~\ref{tab:exp_BD_PSNR}. 

In terms of PSNR-RGB, one can see that our method shows comparable performance to the state-of-the-art learned codecs, CVPR'20~\cite{ContextGMM} and TPAMI'20~\cite{iwave}, on Kodak and CLIC datasets. Remarkably, it achieves the best performance among all the learned codecs on Tecnick and CLIC datasets. It however falls short of the VTM model slightly on Kodak, Tecnick and CLIC datasets. In particular, ANFIC displays a tendency to perform worse at low rates. This may be attributed to the fact that additive coupling layers are susceptible to the accumulation and propagation of quantization errors (Fig.~\ref{fig:error}). It is important to note in Table~\ref{tab:exp_BD_PSNR} that ANFIC is inferior to VTM in BD-rate saving by a significant margin ($7\%$) on CLIC Professional dataset. Careful examination of the dataset reveals that some images are extremely challenging and not typical of the images found in our training data. All the competing methods are faced with the same issue. It is expected that increasing the diversity of training data will help. Nevertheless, the superiority of ANFIC over BPG is apparent on all the datasets.  

In terms of MS-SSIM, our method performs among top two. It is slightly worse than the top performer, CVPR'20~\cite{ContextGMM}, on Kodak dataset, especially at low rates (See Fig.~\ref{fig:KodakSSIM}), but is comparable to ICLR'19~\cite{lee}, which achieves the best MS-SSIM performance on the CLIC dataset. It is worth mentioning that TPAMI'20~\cite{iwave}, a strong baseline when evaluated with PSNR-RGB, exhibits poor MS-SSIM results because the released model is optimized for MSE only. Also, as noted previously in other studies, the learned codecs outperform VTM and BPG considerably when trained and tested by MS-SSIM.  

The model size comparison in Table~\ref{tab:exp_BD_PSNR} suggests that the rate-distortion benefits of ANFIC do not come at the expense of unreasonably huge models. Its model size is between that of TPAMI'20~\cite{iwave} and CVPR'20~\cite{ContextGMM}, both show competitive rate-distortion performance.  

\begin{table}[t]
\centering%
\caption{Comparison of the BD-rate savings and model sizes of the competing methods (optimized by MSE). The BD-rate savings are reported with BPG-444 serving as the anchor. The best performer is marked with ``\textcolor{red}{$^\dagger$}", and the second best with ``\textcolor{blue}{$^\ast$}".}
\label{tab:exp_BD_PSNR}
\setlength{\tabcolsep}{3mm}{
\scalebox{0.84}{
    \begin{tabular}{|c|c|c|c|c|c|}
    \hline
    \multirow{2}{*}{\textbf{Methods}} &
    \multicolumn{4}{c|}{\textbf{BD-rate (\%)}} &
    \multirow{2}{*}{\textbf{Model Size}}\\
    \cline{2-5}
    \multirow{1}{*}{} &
    \multicolumn{1}{c|}{\textbf{Kodak}} &
    \multicolumn{1}{c|}{\textbf{Tecnick}} &
    \multicolumn{1}{c|}{\textbf{CLIC}} &
    \multicolumn{1}{c|}{\textbf{CLIC Pro}}&
    \multirow{1}{*}{}\\
    
    \hline
    ICLR'18~\cite{GoogleHyperPrior}& 5.0 & 2.9 & 13.2 & -   & 12M \\
    \hline
    NIPS'18~\cite{ContextModel}& -3.0 & -15.6 & 0.3 & -   & 20M \\
    \hline
    ICLR'19~\cite{lee}& -3.0 & -10.1 & -7.5 & -13.8 & 73M \\
    \hline
    TPAMI'20~\cite{iwave} & \textcolor{blue}{-16.3$^\ast$} & -22.6  & -18.1  & -20.3  & 18M \\
    \hline
    CVPR'20~\cite{ContextGMM}& -14.5  & -   & -   & \textcolor{blue}{\textcolor{blue}{-25.3$^\ast$}} & 27M \\
    \hline
    TPAMI'21~\cite{benchmark}   & -8.8 & -15.0 & -13.0  & -   & 73M\\
    \hline
    ANFIC (Ours)  & -15.3  & \textcolor{blue}{-26.5$^\ast$}  & \textcolor{blue}{-18.5$^\ast$}  & -24.5  & 23M\\
    \hline
    VTM-444   & \textcolor{red}{-17.9$^\dagger$} & \textcolor{red}{-29.7$^\dagger$} & \textcolor{red}{-22.6$^\dagger$} & \textcolor{red}{-31.3$^\dagger$} & - \\
    \hline
    \end{tabular}
}
}
\end{table}
\vspace{-0.4em}

\subsection{Subjective Quality Comparison}
\label{subsec:subjective}

\begin{figure*}[t!] %!t
\centering
\includegraphics[width=0.85\linewidth]{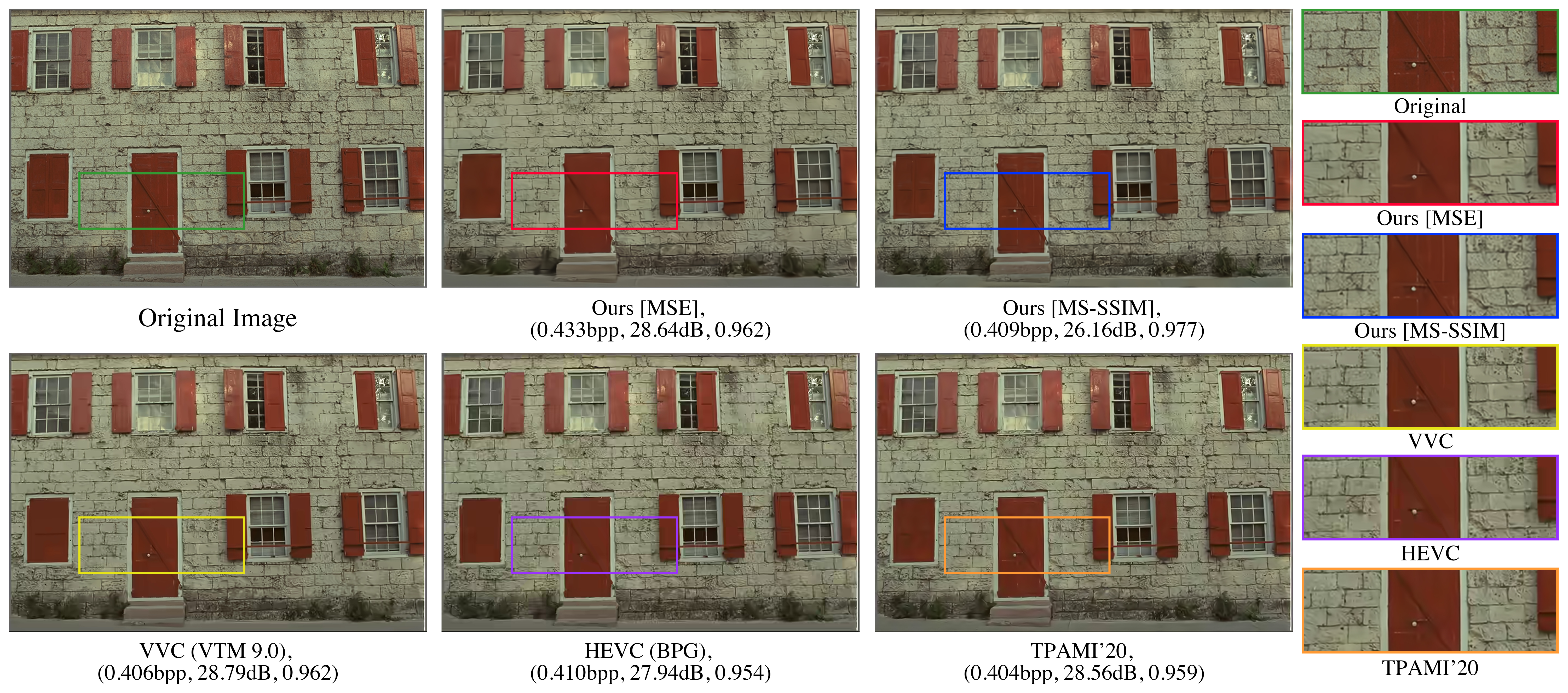}
\caption{Subjective quality comparison of image $kodim01$ from Kodak dataset.}
\label{fig:subjQ01}
%\vspace{-2.0em}
\end{figure*}

\begin{figure*}[t!] %!t
\centering
\includegraphics[width=0.85\linewidth]{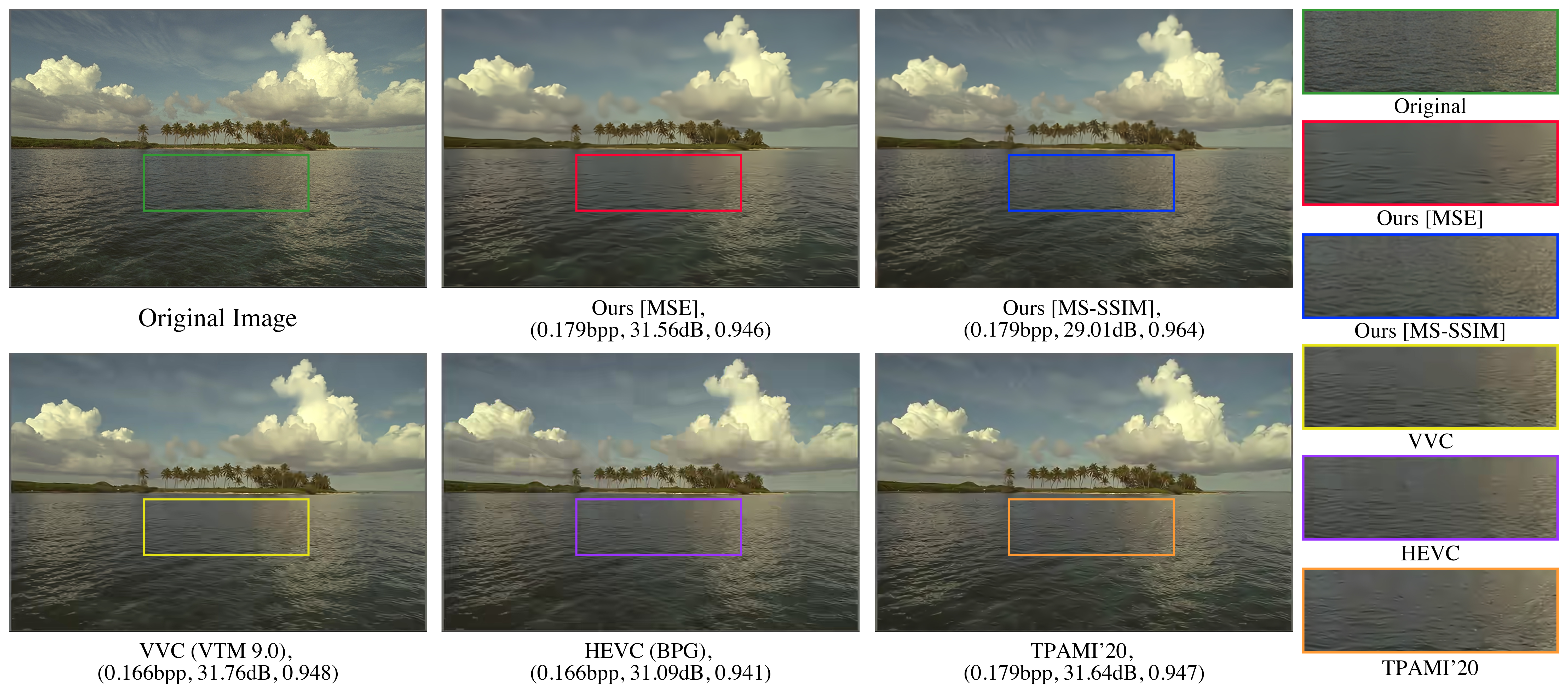}
\caption{Subjective quality comparison of image $kodim16$ from Kodak dataset.}
\label{fig:subjQ16}
\vspace{-1.0em}
\end{figure*}

%\vspace{-0.6em}
\begin{comment}
\begin{figure*}[t!] %!t
\centering
\includegraphics[width=0.9\linewidth]{FIG/SubjectiveQuality.jpg}
\caption{Subjective quality comparison of image $kodim21$ from Kodak dataset.}
\label{fig:subjQ21}
% %\vspace{-2.0em}
\end{figure*}
\end{comment}

%\begin{figure*}[t!] %!t
%\centering
%\includegraphics[width=0.9\linewidth]{FIG/SubjectiveQuality4.png}

%\caption{Subjective quality comparison of images $carrots$ from Tecnick dataset.}
%\label{fig:subjQca}
%\vspace{-2.0em}
%\end{figure*}

%\begin{figure*}[t!] %!t
%\centering
%\includegraphics[width=0.9\linewidth]{FIG/SubjectiveQuality5.png}

%\caption{Subjective quality comparison of images $tomatos_a$ from Tecnick dataset.}
%\label{fig:subjQto}
%\vspace{-2.0em}
%\end{figure*}

Figs.~\ref{fig:subjQ01} and ~\ref{fig:subjQ16} show the subjective quality comparison between ANFIC (ours), VVC, BPG, and TPAMI'20~\cite{iwave} on images $kodim01$ and $kodim16$ from Kodak dataset. It is seen that our MSE model achieves comparable subjective quality to VVC and TPAMI'20~\cite{iwave}. As expected, ANFIC optimized for MSE tends to smooth the highly-textured areas, while VVC and HEVC generates clear blocking 
artifacts in Fig.~\ref{fig:subjQ16}. In particular, TPAMI'20~\cite{iwave} suffers from geometric distortion especially in the "door" area in Fig.~\ref{fig:subjQ01} and produces some artificial noisy dots on the "water surface" in Fig.~\ref{fig:subjQ16}. 
In contrast, our MS-SSIM model shows much better subjective quality, preserving most high-frequency details.

%Since the performance of our ANFIC could degrade due to the training volatility when the architecture of ANFIC becomes deeper, we simplify our autoencodeing transform with only the addition operation to improve the training stability, treating $s_\pi^{enc}(x)$ and $\sigma_\pi^{dec}(z)$ as 1 in Eq.~\eqref{eq:ANF_2}.

%Since our ANFIC can offer a wide range of quality levels, we use it as anchor in reporting the BD-rates of the  baseline models, some of which work better at lower rates while the others are optimized for higher rates. Note that rate inflation as compared to ANFIC is reflected by positive BD-rates while rate saving is shown as negative BD-rates. 
%The resolution ranges from $1803 \times 1175$ for professional to $1913 \times 1361$ for mobile.
\vspace{-0.6em}
\section{Ablation Studies}
\label{sec:ablation}

In this section, we conduct ablation studies to understand ANFIC's properties. Firstly, we show how the ANF framework improves the VAE-based scheme by stacking its autoencoding transform (Section~\ref{subsec:effectiveness}). Secondly, we investigate the effect of the quality enhancement network on ANFIC and its VAE-based counterpart (Section~\ref{subsec:quality_enhancement}). Thirdly, we discuss the effect of imposing different regularization strategies on $x_2$ (Section~\ref{subsec:x2}). Fourthly, we analyze the inner workings of ANFIC by visualizing the output of each autoencoding transform in both spatial and frequency domains (Section~\ref{subsec:step-by-step}). Fifthly, we study the compression performance of ANFIC across low and high rates (Section \ref{subsec:nl}). Finally, we extend ANFIC to support variable rate compression and compare its performance with the other baselines (Section \ref{subsec:vr}). Unless otherwise specified, Kodak dataset is used for ablation experiments.

\begin{figure}[t!] %!t
\centering
\includegraphics[width=0.8\linewidth]{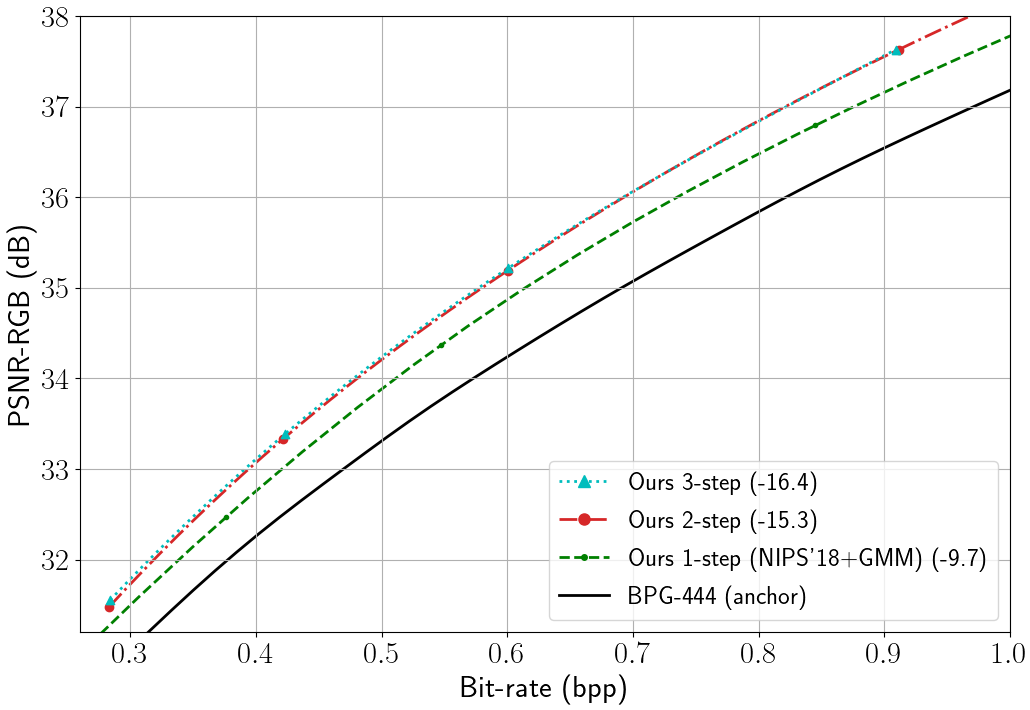}
\caption{Rate-distortion curves for different number of autoencoding transforms.}
\label{fig:ab_1}
\vspace{-1em}
\end{figure}

\begin{figure}[t!] %!t
\centering
\includegraphics[width=0.8\linewidth]{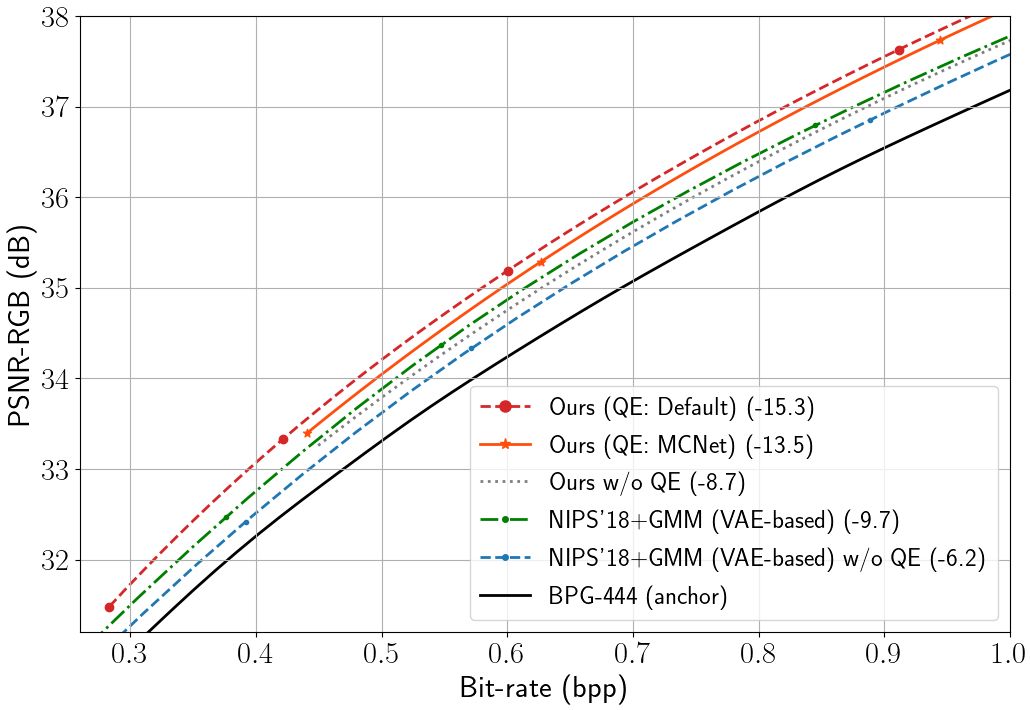}
\caption{Rate-distortion performance with and without the quality enhancement network.}
\label{fig:ab_2}
\vspace{-1em}
\end{figure}

\begin{comment}
\begin{figure*}[t!]
\begin{center}
\begin{subfigure}{0.48\textwidth}
    % include first image
    \centering
    %\vspace{-0.8em}
    \includegraphics[width=\linewidth]{FIG/crop_R_D_ArchiCmp.png} 
    \caption{}
    %\vspace{-1.2em}
    \label{fig:ab_1}
\end{subfigure}
\begin{subfigure}{0.48\textwidth}
    % include first image
    \centering
    %\vspace{-0.8em}
    \includegraphics[width=\linewidth]{FIG/crop_R_D_ArchiCmp.png}
    \caption{}
    %\vspace{-1.2em}
    \label{fig:ab_2}
\end{subfigure}

\caption{Rate-distortion curves for different number of autoencoding transforms. Rate-distortion performance with and without the quality enhancement network.}
\label{fig:effectiveness}
\end{center}
%\vspace{-2.0em}
\end{figure*}
\end{comment}

\subsection{Number of Autoencoding Transforms}
\label{subsec:effectiveness}

To see the rate-distortion benefits of stacking autoencoding transforms, we compare between the VAE-based scheme~\cite{ContextModel} and ANFIC with a varied number of autoencoding transforms. It is important to note that the VAE-based scheme can be interpreted as one-step ANFIC (see Section~\ref{subsec:ANFIC}). For a fair comparison, the VAE-based scheme (which is termed ``NIPS'18+GMM" and is modified from~\cite{ContextModel} by additionally including Gaussian mixture-based entropy coding and the quality enhancement network~\cite{iwave}) and ANFIC share the same autoencoding architecture, entropy coding scheme, and quality enhancement network. To keep the model size comparable, the channel number of every autoencoding transform in ANFIC is set to 128 (See Fig. \ref{fig:architecture}), while that of the VAE-based counterpart is 192. This ensures that ANFIC with two autoencoding transforms (the main setting used throughout this paper) has a similar model size to the VAE-based one. Nevertheless, when the number of autoencoding transforms increases beyond two, the model size of ANFIC increases linearly. 

From Fig.~\ref{fig:ab_1}, it is seen that increasing the number of autoencoding transforms from one layer (VAE-based) to two layers (Ours 2-step) improves the rate-distortion performance significantly. However, the gain diminishes sharply when the number goes beyond two. We thus choose two autoencoding transforms as our default setting. 

A side experiment shows that increasing the channel number (i.e. the L value in Fig.~\ref{fig:architecture}) of the autoencoding transform from 128 to 192 improves the BD-rate saving only marginally by 1.1\%. The channel number is defaulted to 128 for lower complexity and fair comparison.

\subsection{Effect of Quality Enhancement Networks}
\label{subsec:quality_enhancement}
Fig.~\ref{fig:ab_2} shows the effect of the quality enhancement network (as a post-processing network) on the rate-distortion performance of ANFIC and the VAE-based scheme~\cite{ContextModel}. In addition to the default quality enhancement network from~\cite{iwave}, we experiment with another popular one, known as MCNet~\cite{dvc}, which is often used in the end-to-end learned video codecs to enhance the quality of the motion-compensated frame~\cite{dvc}. The two quality enhancement networks have similar model sizes. The major difference between them is that the default one~\cite{iwave} does not have striding and pooling operations, whereas MCNet~\cite{dvc} has a U-net structure, where the resolution of the feature maps shrinks first and stretches later.

We observe that ANFIC benefits more from the use of the default quality enhancement network~\cite{iwave}, which boosts the BD-rate saving of ANFIC by $6.6\%$ as compared to $3.5\%$ with the NIPS'18+GMM (VAE-based with default quality enhancement network~\cite{iwave}) scheme~\cite{ContextModel}. This suggests that ANFIC literally separates the image transformation and the (quantization) error compensation into two orthogonal parts. The former is addressed by invertible autoencoding transforms while the latter relies on the quality enhancement network. The fact that the feature extraction and the image reconstruction in ANFIC have to go through the same invertible coupling layers make it difficult to learn autoencoding transforms that can handle well both image representation and error compensation. This however is not the case with the NIPS'18+GMM (VAE-based) scheme, where the analysis and the synthesis transforms do not share the same network. Usually, the synthesis transform can learn to compensate partially for quantization errors. As such, the gain from the quality enhancement network becomes limited when the synthesis network is already capable enough. %The result is in line with an earlier finding reported in~\cite{iwave} that the flow-based model can benefit more from post-processing.

From Fig.~\ref{fig:ab_2}, it is also seen that the default quality enhancement network~\cite{iwave} shows better rate-distortion performance than MCNet~\cite{dvc}, especially at lower rates. This may be attributed to the fact that the striding and pooling of MCNet~\cite{dvc} could cause the loss of some spatial information. In any case, ANFIC with either quality enhancement network outperforms NIPS’18+GMM.

\subsection{Effect of $x_2$ Regularization}
\label{subsec:x2}

\begin{figure}[t!] %!t
\centering
\includegraphics[width=0.8\linewidth]{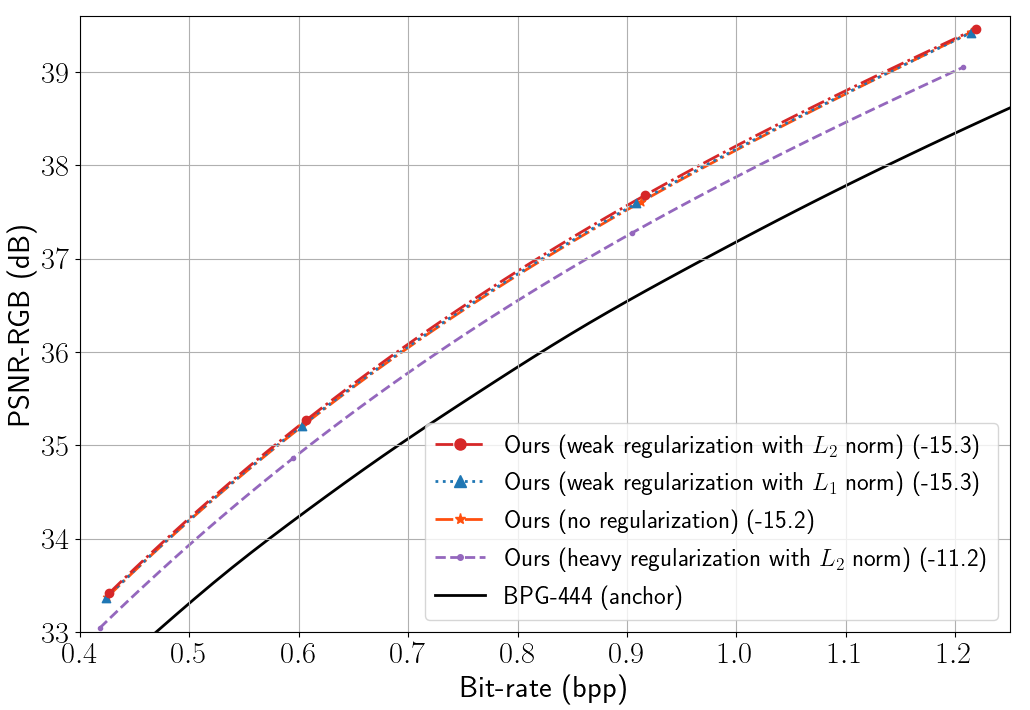}
\caption{ANFIC with different regularization strategies imposed on $x_2$.}
\label{fig:x2}
%\vspace{-0.3em}
\end{figure}

Fig.~\ref{fig:x2} compares the rate-distortion curves for different regularization strategies imposed on $x_2$, including (1) weak regularization ($\lambda_1 = 0.01 * \lambda_2$) with the $L_2$ norm (the proposed method), (2) weak regularization ($\lambda_1 = 0.01 * \lambda_2$) with the $L_1$ norm, (3) heavy regularization ($\lambda_1 = 1 * \lambda_2$) with the $L_2$ norm, and (4) no regularization ($\lambda_1 = 0$). It can be observed that weak regularization with either the $L_2$ norm or $L_1$ norm achieves the best rate-distortion performance, presenting 15.3\% BD-rate reductions. Heavy regularization with the $L_2$ norm, however, degrades the rate-distortion performance, because the regularization loss is weighted equally as the reconstruction loss. No regularization, interestingly, shows marginally worse rate-distortion performance (15.2\% BD-rate reduction) than the weak regularization with the $L_2$ norm (the proposed method).

The fact that no regularization shows marginal impact on the final rate-distortion performance has partially to do with our setting $x_2$ to 0 for reconstruction during training. Recall that the mapping between the input $(x, e_z, e_h)$ and the latent representation $(x_2, z_2, h_2)$ is invertible (See Fig.~\ref{fig:overall}). In the absence of quantization, using $(0, z_2, h_2)$ in place of $(x_2, z_2, h_2)$ for decoding while ensuring the invertibility by minimizing the reconstruction loss $d(x, \hat{x})$ would compel $x_2$ to approximate noughts during encoding without any additional regularization. The same trend carries roughly over to the case when $x_2, z_2, h_2$ are quantized. We however notice that imposing weak regularization on $x_2$ during encoding will make the training more stable.

\begin{comment}
\begin{figure*}[t!]
\centering
%\setlength{\tabcolsep}{1pt}
\resizebox{1\linewidth}{!}
{
\Huge
\begin{tabular}{cccc}
    $x$ & $x_1 = x - \mu^{dec}_{\pi_1}$ & $x_2 = x_1 - \mu^{dec}_{\pi_2}$ & $x_3 = x_2 - \mu^{dec}_{\pi_3}$\\
    \includegraphics[width=0.4\textwidth]{FIG/H1.png}&
    \includegraphics[width=0.4\textwidth]{FIG/H3.png}&
    \includegraphics[width=0.4\textwidth]{FIG/H5.png}&
    \includegraphics[width=0.4\textwidth]{FIG/H7.png}\\
    
    $\mu^{dec}_{\pi_1}$ & $\mu^{dec}_{\pi_2}$ & $\mu^{dec}_{\pi_3}$ &  $\hat{x}$\\
    \includegraphics[width=0.4\textwidth]{FIG/H2.png}&
    \includegraphics[width=0.4\textwidth]{FIG/H4.png}&
    \includegraphics[width=0.4\textwidth]{FIG/H6.png}&
    \includegraphics[width=0.4\textwidth]{FIG/H8.png}\\
\end{tabular}
}
\caption{Visualization of our ANF transformation in high rate.}
%\vspace{-0.3cm}
\label{fig:evolution_H}
\end{figure*}
\end{comment}

\begin{figure}[t!]
%\vspace{-0.5cm}
\centering
\resizebox{1\linewidth}{!}
{
\Huge
\begin{tabular}{cccc}
    $x$ & $x_1 = x - \mu^{dec}_{\pi_1}$ & $x_2 = x_1 - \mu^{dec}_{\pi_2}$ & $x_3 = x_2 - \mu^{dec}_{\pi_3}$\\
    \includegraphics[width=0.35\textwidth]{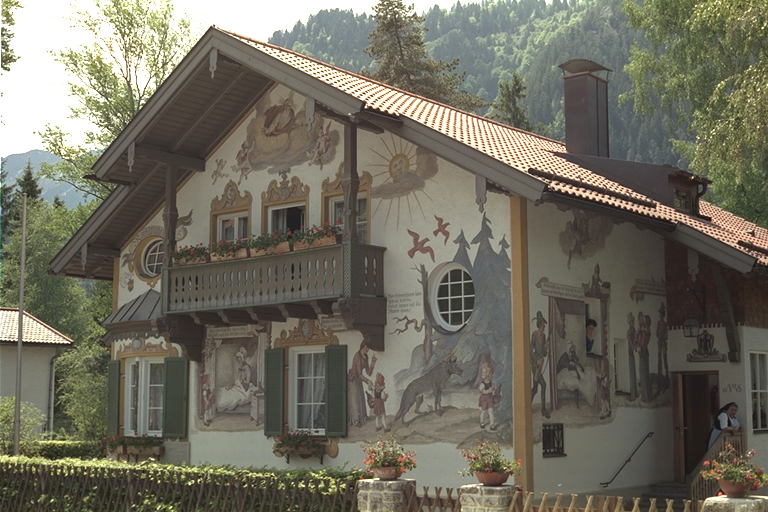}&
    \includegraphics[width=0.35\textwidth]{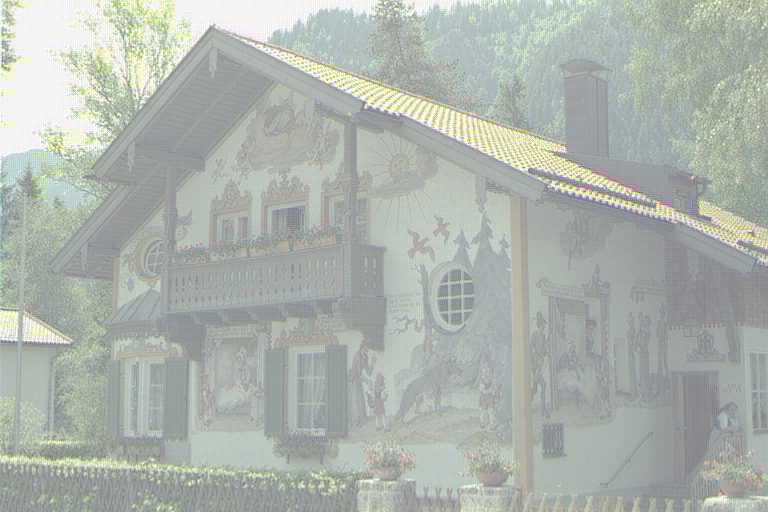}&
    \includegraphics[width=0.35\textwidth]{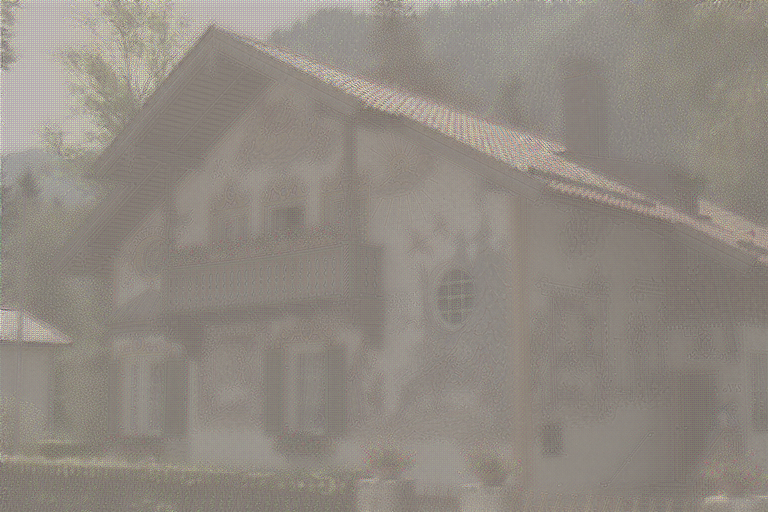}&
    \includegraphics[width=0.35\textwidth]{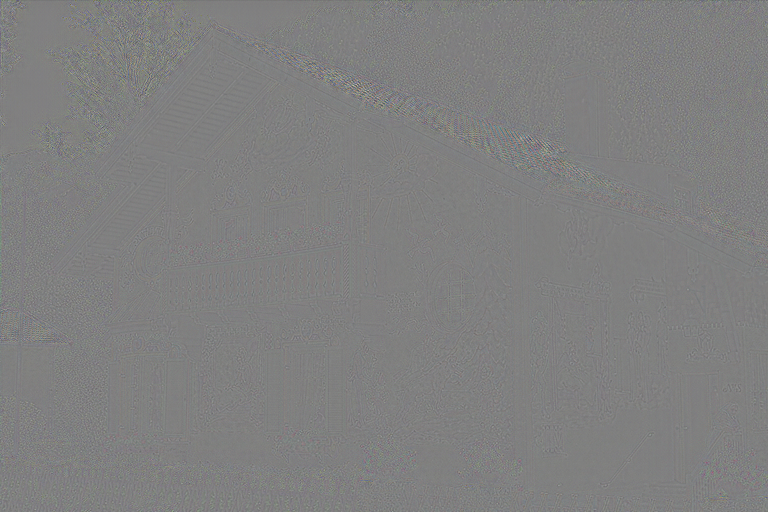}\\
    \includegraphics[width=0.35\textwidth]{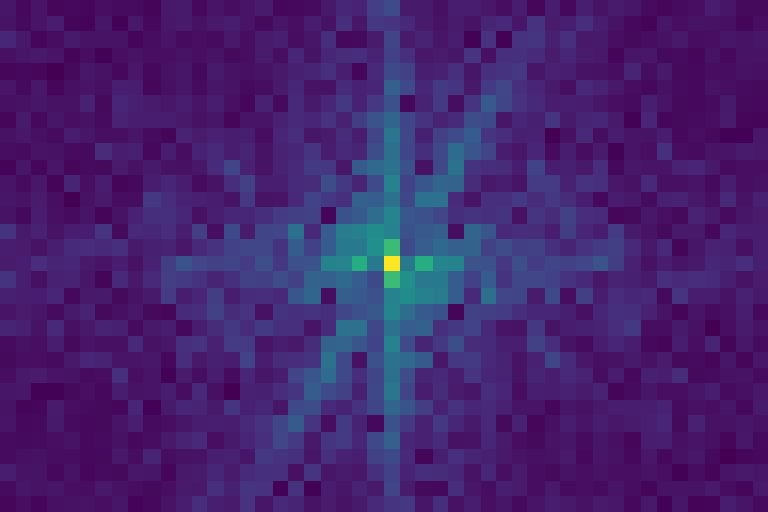}&
    \includegraphics[width=0.35\textwidth]{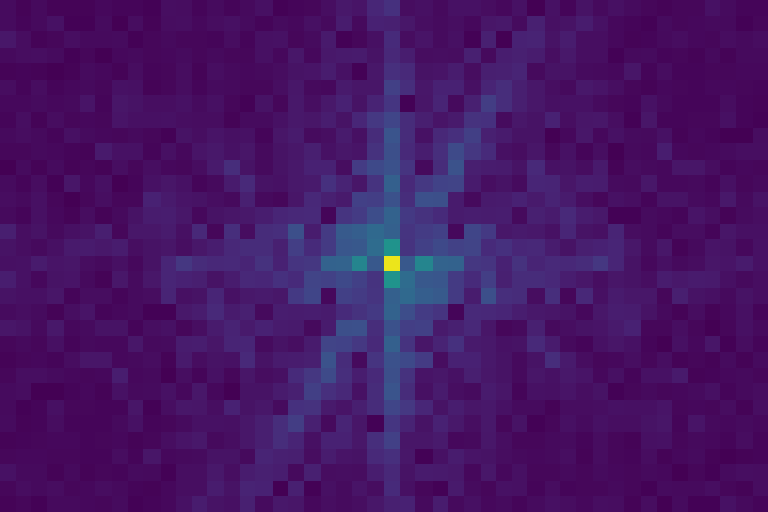}&
    \includegraphics[width=0.35\textwidth]{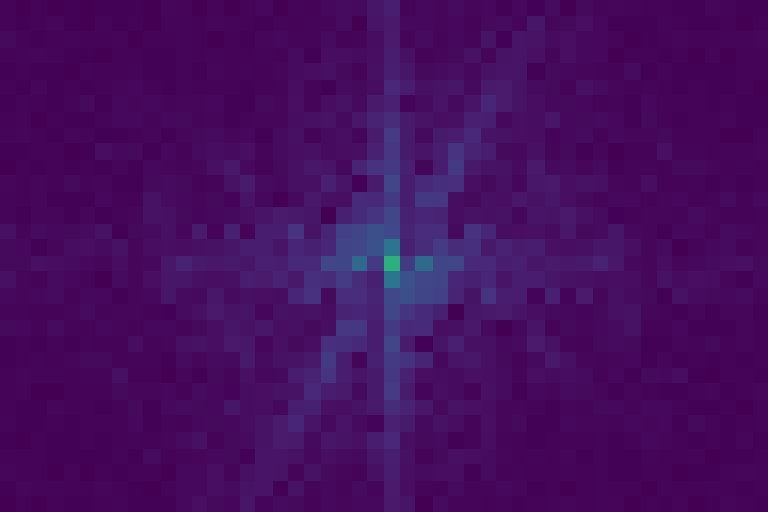}&
    \includegraphics[width=0.35\textwidth]{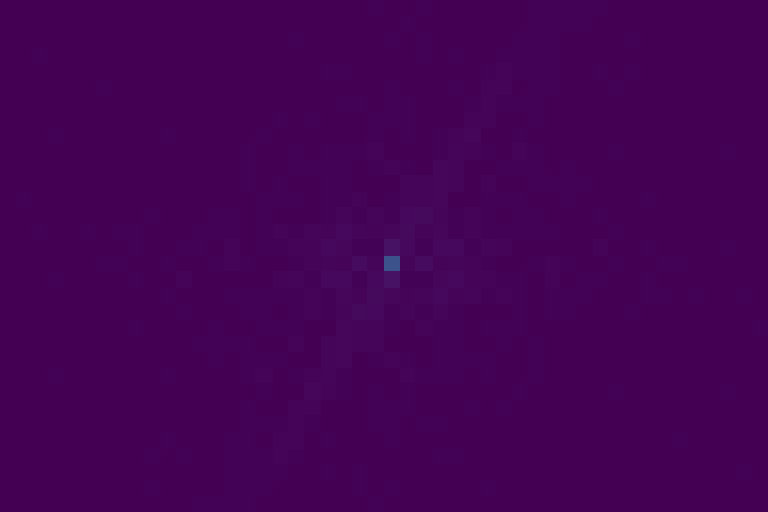}\\
    $\mu^{dec}_{\pi_1}$ & $\mu^{dec}_{\pi_2}$ & $\mu^{dec}_{\pi_3}$ &  $\hat{x}$\\
    \includegraphics[width=0.35\textwidth]{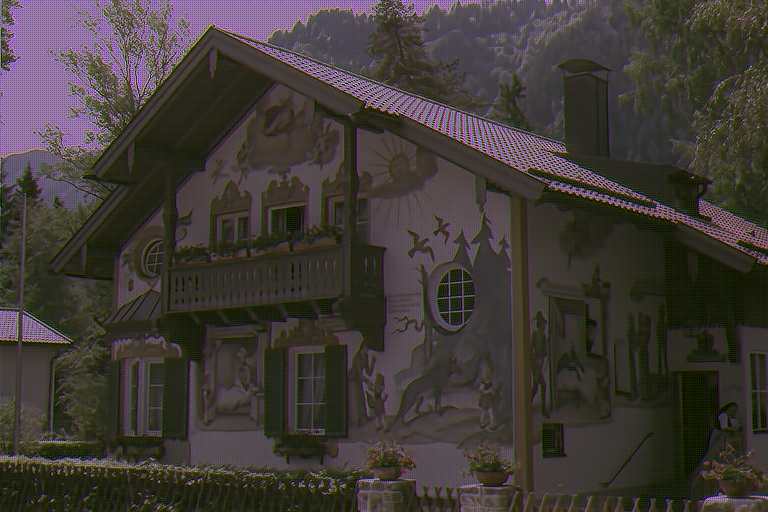}&
    \includegraphics[width=0.35\textwidth]{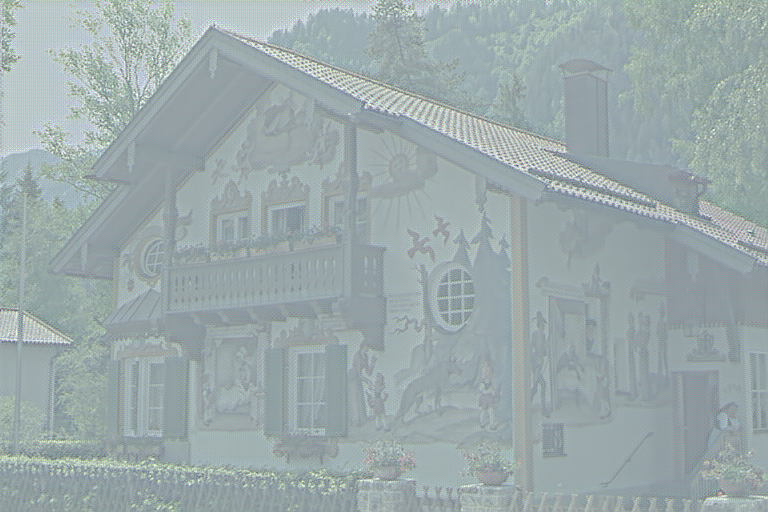}&
    \includegraphics[width=0.35\textwidth]{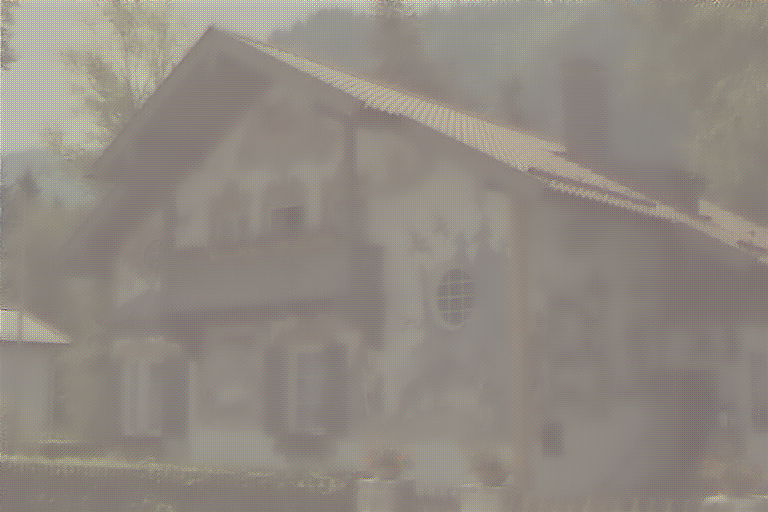}&
    \includegraphics[width=0.35\textwidth]{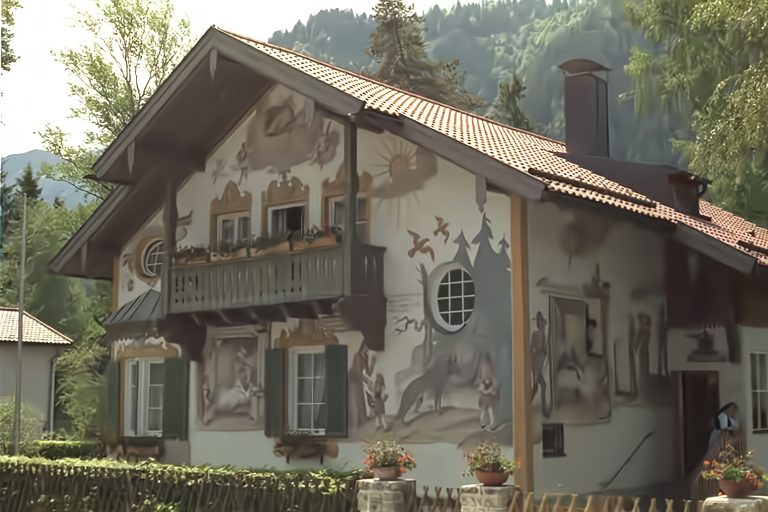}\\
    \includegraphics[width=0.35\textwidth]{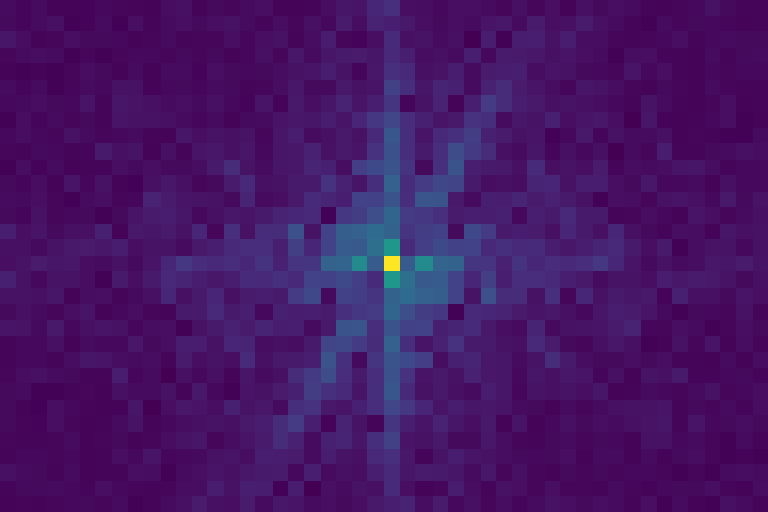}&
    \includegraphics[width=0.35\textwidth]{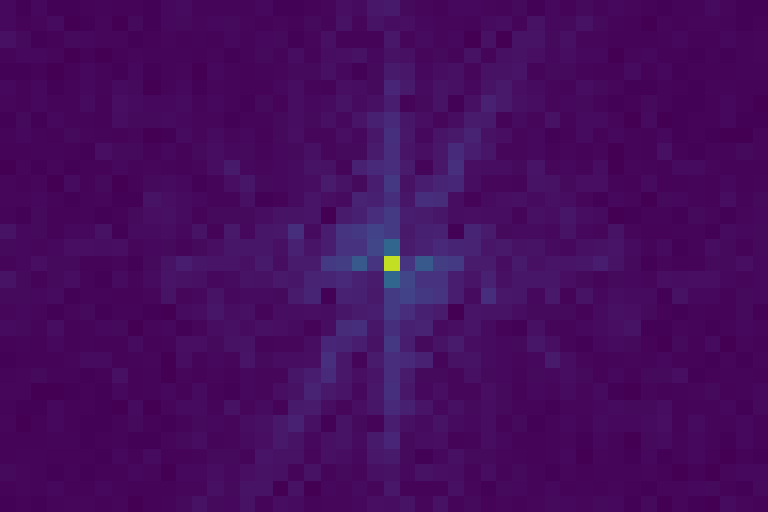}&
    \includegraphics[width=0.35\textwidth]{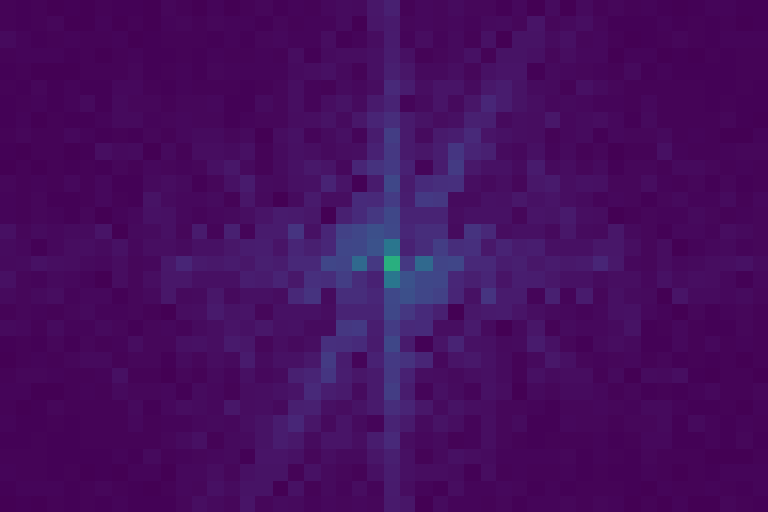}&
    \includegraphics[width=0.35\textwidth]{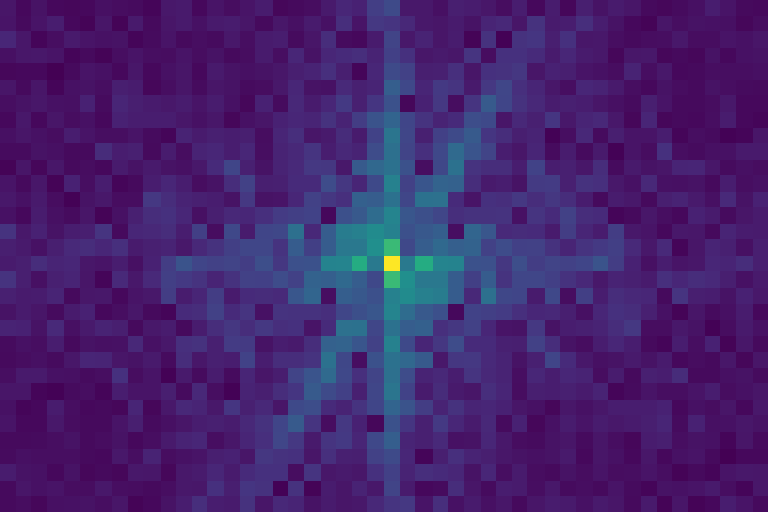}\\
\end{tabular}
}
\caption{Visualization of the autoencoding transform outputs $\{x_i\}_{i=1}^3$ and the decoder outputs $\{\mu_{\pi_i}^{dec}\}_{i=1}^3$ in the autoencoding transforms in three-step ANFIC, where the average image intensity has been shifted to 128 for better viewing. The signal spectra in frequency domain are plotted as heatmaps.}
%\vspace{-0.3cm}
\label{fig:evolution_L}
\end{figure}

\vspace{-0.6em}
\begin{figure}[t!] %!t
\centering
\includegraphics[width=0.95\linewidth]{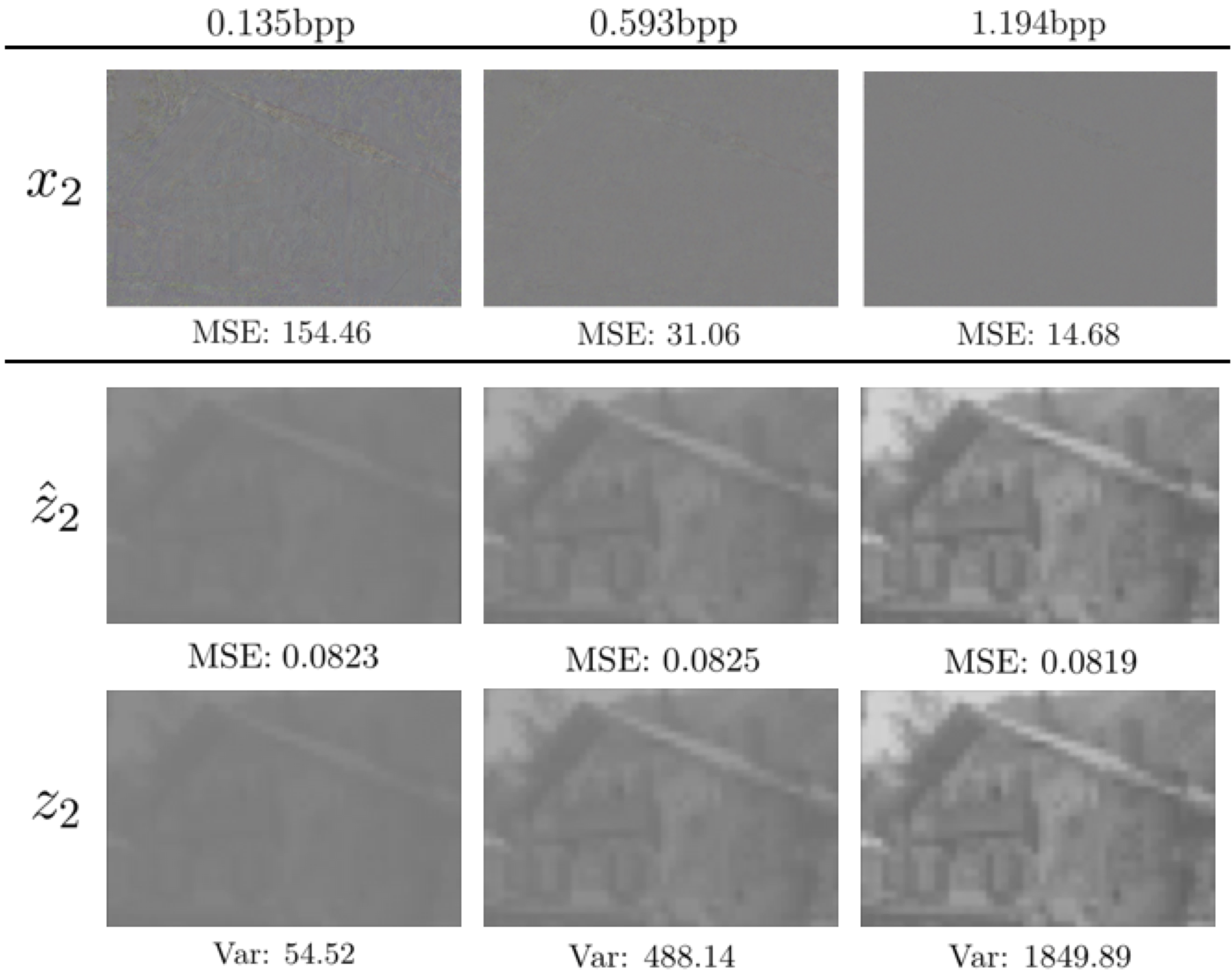}
\caption{Visualization of $x_2$, $\hat{z}_2$, and $z_2$, where only few channels with the largest variances are shown for $\hat{z}_2$ and $z_2$. The Mean Square Error (MSE) of $x_2$ is measured against a zero image while that of $\hat{z}_2$ is against $z_2$.}
\label{fig:x2z2}
\vspace{-1em}
\end{figure}

\vspace{-0.6em}
\subsection{Visualization of Autoencoding Transforms}
\label{subsec:step-by-step}

Fig.~\ref{fig:evolution_L} visualizes how our ANFIC model (see Fig.~\ref{fig:architecture}) transforms the input image $x$ step-by-step into a residual image $x_2$ and what information is captured by the corresponding latent code $z_i,i=1,2$ in each step. Additionally, the corresponding signal spectra in frequency domain are presented to understand the system response of every autoencoding transform. For better visualizing the evolution of signals, we extend the architecture in Fig.~\ref{fig:architecture} to three-step ANFIC, with the final outputs being $x_3$ and $z_3$ (instead of $x_2$ and $z_2$ as depicted in Fig.~\ref{fig:architecture}). Also presented in this figure are the decoder outputs $\{\mu^{dec}_{\pi_i}\}_{i=1}^3$ of the autoencoding transforms (see Eq.~\eqref{eq:ANF_2} and Fig.~\ref{fig:architecture}), which reveal the information captured by the latent code $\{z_i\}_{i=1}^3$. As an example, the first autoencoding transform converts the image $x$ into the latent code $z_1$, which is then decoded as $\mu^{dec}_{\pi_1}$ to be subtracted from $x$. Hence, $\mu^{dec}_{\pi_1}$ stands for an estimate of $x$ that is derived from the latent $z_1$. 

From left to right in the top two rows, one can see that the high-frequency details of the input image $x$ are filtered out in successive autoencoding transforms, arriving at a residual image $x_3$ with little high-frequency information (see the sub-figure in the top-right corner). As such, the autoencoding transforms in ANFIC act as low-pass filters, where their cut-off frequency decreases with the increasing transform step in the feature extraction process. Because $x_3$ will be discarded during the reconstruction process, the remaining high-frequency details in $x_3$ will be lost completely. Thus, ANFIC is lossy.

The decoder outputs of the autoencoding transforms further shed light on how the latent code is transformed from $e_z$ into a form suitable for compression (i.e. $e_z \rightarrow z_1 \rightarrow z_2 \rightarrow z_3$). From left to right in the bottom two rows, we see that $\mu^{dec}_{\pi_1}$ (decoded from $z_1$) presents a rough estimate of the input $x$. Its spectrum looks similar to that of $x$, but is not exactly the same. We conjecture that $\mu^{dec}_{\pi_1}$ focuses more on the approximation of the high-frequency part of the input $x$. The corroborating fact is that when it is subtracted from $x$, the resulting output $x_1=x-\mu^{dec}_{\pi_1}$ has relatively less high-frequency information. This becomes even more obvious in the following autoencoding transform, where $\mu^{dec}_{\pi_2}$ (decoded from $z_2$) addresses primarily the remaining mid-frequency part in $x_1$; as a result, the output $x_2=x_1-\mu^{dec}_{\pi_2}$ of the second transform becomes an even  lower-frequency signal. In the end, the latent code $z_3$, which will be compressed into the bitstream, only needs to represent a low-pass filtered version of the original input, which is relatively easy to compress. The reconstruction process updates a zero image in $x_3$ by those decoder outputs in reverse order (i.e. $\mu^{dec}_{\pi_3} \rightarrow \mu^{dec}_{\pi_2} \rightarrow \mu^{dec}_{\pi_1}
\rightarrow x$), to recover the low-frequency, mid-frequency, and high-frequency details of the input $x$ step-by-step.  

Fig.~\ref{fig:x2z2} further visualizes $x_2$, $z_2$, and $\hat{z}_2$ at different bit rates ranging from 0.135bpp to 1.194bpp. It is seen that more residuals appear in $x_2$ at low rates than at high rates, suggesting that setting $x_2$ to a zero image at low rates would introduce more distortion than at high rates. As for $z_2$ and $\hat{z}_2$, because a fixed, uniform quantization step size, i.e. 1, is used for all the rate points, the MSE between $z_2$ and $\hat{z}_2$ does not change significantly. However, the network learns to adjust the variance of $z_2$ in order to control the signal-to-noise ratio in the latent space. We see that the higher the bit rate is, the more information is captured by $z_2$; namely, $z_2$ tends to have larger variances at high rates. All in all, the information captured by $x_2$ decreases with the increasing bit rate, whereas that by $z_2$ increases accordingly.

%From the FFT result, $\mu^{dec}_{\pi_1}$ contains some high-frequency information and it looks similar but not exactly the same as the frequency response of $x$. After transforming $x$ by subtracting the $\mu^{dec}_{\pi_1}$, $x_1$ becomes a blurry image which contains relatively low-frequency information compared with $x$. 

%The similar-behaved second autoencoding transformation transforms the $x_1$ into a more blurry image $x_2$ with even lower frequency response in its content. The ANFIC repeats the third autoencoding transformation and finally reaches to a residual image $x_3$ with few super high-frequency information that is non-transformable under the given bitrate constraint. Since this super high-frequency information will be discarded during the decoding process, $\hat{x}$ can not reconstruct those area such as the stripped pattern in the red box. In general, each layer of ANFIC operates as a low-pass filter to remove the high-frequency information from the previous $x_i$, and yielding a nearly zero image as $x_3$.

\begin{figure}[t!] %!t
\centering
\includegraphics[width=0.8\linewidth]{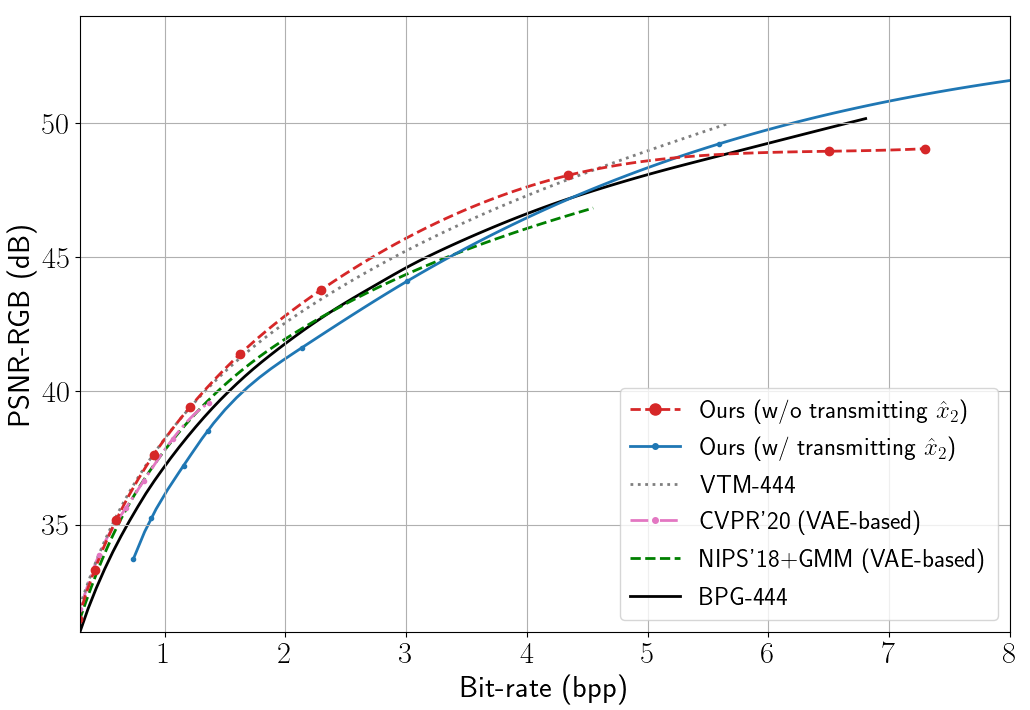}
\caption{Rate-distortion comparison between our ANFIC and VAE-based schemes across low and high rates.}
\label{fig:NLL}
\vspace{-0.3em}
\end{figure}

\subsection{Compression Performance across Low and High Rates}
\label{subsec:nl}
This study investigates the compression performance of ANFIC over a wide range of bit rates. It is reported in~\cite{NFcodec, iwave} that most VAE-based compression schemes suffer from the autoencoder limitation; that is, the reconstruction by the autoencder is generally lossy, even without quantization. As a result, it is difficult for a VAE-based model to support efficient compression over a wide rage of bit rates without changing the network architecture, for example, by adjusting the number of channels. ANFIC, although being a flow-based model, is lossy due to discarding the high-frequency information in the residual image $x_2$ (see Fig.~\ref{fig:architecture}) for reconstruction. 

Fig.~\ref{fig:NLL} compares ANFIC with two state-of-the-art VAE-based schemes over a wide range of bit rates. In particular, ANFIC has the same number of channels (i.e. 320 channels) in latent space as NIPS'18~\cite{ContextModel}, whereas  CVPR'20~\cite{ContextGMM} has only 192 channels yet with a larger model size. %The network architectures of all the competing models are fixed except that their network weights are trained separately for different rate points. 
We see that our ANFIC (w/o transmitting $\hat{x}_2$) matches the performance of VTM closely from extremely low-rate compression up to perceptually lossless compression, while the two VAE-based schemes tend to fall short of VTM and even BPG at high rates. The reason why ANFIC is able to work well across low and high rates are two-fold: (1) the ANF-based backbone is fully invertible, and (2) our training strategies, which require $x_2$ to approximate noughts in the feature extraction process and use noughts exactly for $x_2$ during reconstruction, force the image latent $\hat{z}_2$ and its hyperprior $\hat{h}_2$ to capture as much information about the input $x$ as possible (see Fig.~\ref{fig:architecture}).

To further study the invertibility of ANFIC by additionally encoding $x_2$, we model the distribution of the quantized $x_2$, denoted by $\hat{x}_2 = \lfloor x_1 - \mu^{dec}_{\pi_2}(\hat{z}_2) \rceil$, by the convolution of a Gaussian and a uniform distribution. For better coding efficiency, the distribution is conditional on $\hat{z}_2$:
\begin{equation}
    p(\hat{x}_2|\hat{z}_2) = \mathcal N(0, \sigma^{dec}_{\pi_2}(\hat{z}_2)^2) \ast \mathcal{U}(-0.5,0.5)
\end{equation} 
A closer look at the rate-distortion performance w/o and w/ transmitting $\hat{x}_2$ in Fig.~\ref{fig:NLL} reveals that (1) at lower rates, transmitting $\hat{x}_2$ shows worse rate-distortion performance than not transmitting $\hat{x}_2$, and that (2) at higher rates, transmitting $\hat{x}_2$ helps mitigate the quality gap between lossy and (mathematically) lossless compression. In particular, not transmitting $\hat{x}_2$ puts a limit on the highest achievable reconstruction quality (i.e. the rate-distortion curve plateaus after 6bpp). The second observation is in line with the invertibility property of ANF. Focusing on lossy image compression, we opt for not transmitting $\hat{x}_2$ in this paper. However, how to adapt ANFIC to support mathematically lossless coding is an interesting open issue that is among our future work.

%\textcolor{blue}{To further study the invertibility of ANFIC, different from replacing $x_2$ with zero image when decoding, we additionally encoding $x_2$ by introducing a uniform quantization of step size 1 to encode $\hat{x}_2$.}

%\textcolor{blue}{Fig. xx compares the rate-distortion performance between ANFIC w/o transmitting $\hat{x}_2$ and ANFIC w/ transmitting $\hat{x}_2$. We see that the latter shows worse rate-distortion performance. This may be attributed to the fact that $\hat{x}_2$ has the same dimensionality as the input image $x$. Additionally transmitting $\hat{x}_2$ along with the quantized latents $\hat{z}_2$ and $\hat{h}_2$ causes more data to be sent to the decoder. As for high-rate coding, we found that transmitting $\hat{z}_2$ can indeed achieve higher reconstruction quality. This is in line with the invertibility of ANF. That is, the input $x$ can be reconstructed perfectly if both $x_2$ and $z_2$ are sent without quantization and the QE network is identity. This however comes at the cost of a large bit rate. It is to be emphasized that ANFIC is capable of achieving better lossy compression over a wide range of bit rates in terms of rate-distortion trade-offs. How to adapt ANFIC to truly lossless coding is among our future work and is NOT the focus of this paper.}

\vspace{-0.6em}
\begin{figure}[t!] %!t
\centering
\includegraphics[width=0.8\linewidth]{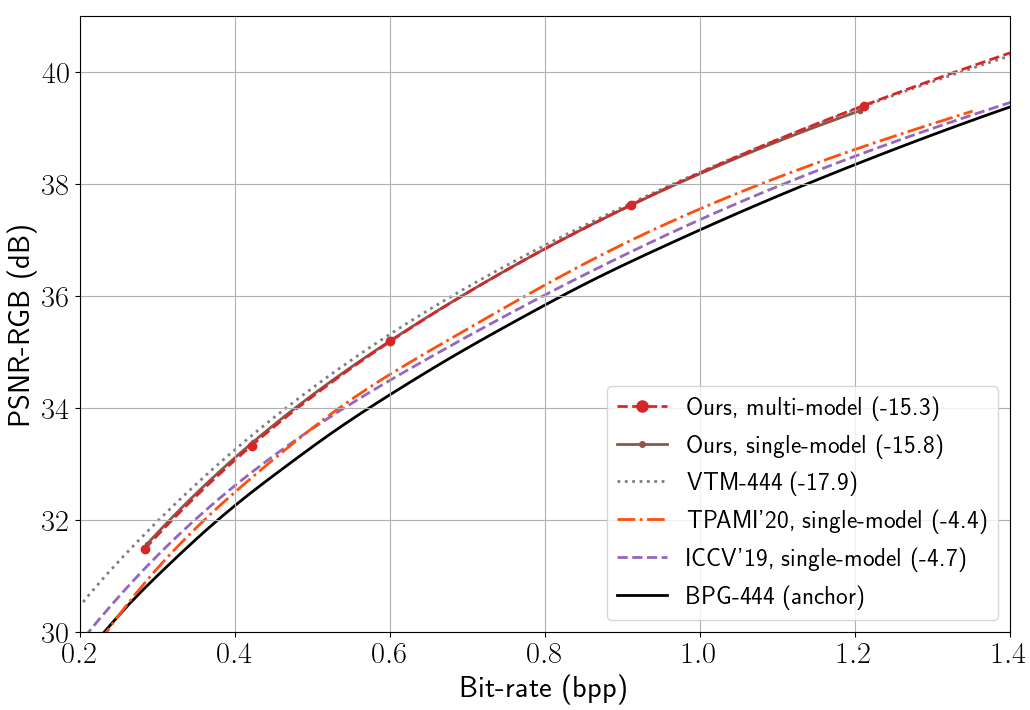}
\caption{Rate-distortion comparison between variable rate models. Multi-model: separate models for distinct rate points; Single-model: a single model for multiple rate points.}
\label{fig:VR}
\vspace{-0.3em}
\end{figure}
\vspace{-0.6em}

\subsection{Variable Rate Compression}
\label{subsec:vr}
Recognizing that ANFIC can work well over a wide range of bit rates, we take one step further to adapt ANFIC to variable rate compression with a single model. To this end, we implement the notion of the conditional convolution in~\cite{conditional}, replacing every convolutional layer with one that is conditional on the $\lambda_2$ (see Eq.~\eqref{eq:training_objective}). The conditional convolution layer applies an affine transformation to every feature map, with the affine parameters derived from a network conditional on the rate parameter $\lambda_2$. For the experiment, we train a single ANFIC model using 5 distinct $\lambda_2$ values $\{0.1, 0.05, 0.02, 0.01, 0.005\}$. The training objective is an extension of Eq.~\eqref{eq:training_objective} by substituting different $\lambda_2$'s into Eq.~\eqref{eq:training_objective} and averaging over these variants. 

Fig.~\ref{fig:VR} shows the rate-distortion comparison of the state-of-the-art variable rate models, including VVC, BPG, ICCV'19~\cite{conditional}, TMAPI'20~\cite{iwave}, and ANFIC (ours). Compared with our multi-model setting, our single-model setting performs comparably well, with slightly increased rate saving due to training variance. It also shows comparable performance to VTM across the 5 rate points, but outperforms significantly the other learning-based methods in single-model mode. 

\section{Conclusion}
\label{sec:conclude}

In this paper, we propose an ANF-based image compression system (ANFIC). It is motivated by the fact that VAE, which forms the basis of most end-to-end learned image compression, is a special case of ANF and can be extended by ANF to offer greater expressiveness. ANFIC is the first work that introduces VAE-based compression in a flow-based framework, enjoying the benefits of both approaches. Experimental results show that ANFIC performs comparably to or better than the state-of-the-art learned image compression and is able to offer a wide range of quality levels without changing the network architecture. Furthermore, its variable rate version shows little performance degradation. Flow-based models are relatively new to learned image compression. We believe there remains widely open space for further research; for example, how to achieve mathematically lossless coding with ANFIC is yet to be addressed.

\vspace{-0.6em}
{\small
\bibliographystyle{ieeetr} % transaction
\bibliography{egbib}
}

\begin{IEEEbiographynophoto}{Yung-Han Ho} received his B.S. and M.S. degrees in electrophysics and electronics engineering from National Chiao Tung University (NCTU) in 2008 and 2010, respectively. 

He is currently pursuing his Ph.D. degree in the department of computer science, National Yang Ming Chiao Tung University (NYCU). His research interests are learning-based image/video coding, computer vision, and machine learning.
\end{IEEEbiographynophoto}
\vspace{-3em}
\begin{IEEEbiographynophoto}{Chih-Chun Chan} received his B.S. degree in computer science and information engineering from National Taiwan Normal University in 2019. 

He is currently pursuing his M.S. degree in the department of computer science, National Yang Ming Chiao Tung University (NYCU). His research interests are learning-based image/video coding, computer vision, and machine learning.
\end{IEEEbiographynophoto}
\vspace{-3em}
\begin{IEEEbiographynophoto}{Wen-Hsiao Peng}(M'09-SM’13) received the B.S., M.S., and Ph.D. degrees from National Chiao Tung University (NCTU), Hsinchu, Taiwan, in 1997, 1999, and 2005, respectively, all in electronics engineering.

He was with the Intel Microprocessor Research Laboratory, Santa Clara, CA, USA, from 2000 to 2001, where he was involved in the development of International Organization for Standardization (ISO) Moving Picture Experts Group (MPEG)-4 fine granularity scalability and demonstrated its application in 3-D peer-to-peer video conferencing. Since 2003, he has actively participated in the ISO/IEC MPEG digital video coding standardization process and contributed to the development of the High Efficiency Video Coding (HEVC) standard and MPEG-4 Part 10 Advanced Video Coding Amd.3 Scalable Video Coding standard. His research group at NCTU is one of the few university teams around the world that participated in the Call-for-Proposals on HEVC and its Screen Content Coding extensions. He is currently a Professor with the Computer Science Department, National Yang Ming Chiao Tung University (NYCU). He was a Visiting Scholar with the IBM Thomas J. Watson Research Center, Yorktown Heights, NY, USA, from 2015 to 2016. He has authored over 70 technical papers in the field of video/image processing and communications and over 60 standards contributions. His research interests include learning-based video/image coding, multimedia analytics, and computer vision.

Dr. Peng is a Technical Committee Member of the Visual Signal Processing and Communications and Multimedia Systems and Application tracks of the IEEE Circuits and Systems Society (CASS). He was Technical Program Co-chair for 2021 IEEE VCIP, 2011 IEEE VCIP, 2017 IEEE ISPACS, and 2018 APSIPA ASC; Publication Chair for 2019 IEEE ICIP; Area Chair/Session Chair/Tutorial Speaker for IEEE ICME and VCIP; and Track Chair/Session Chair/Review Committee Member for IEEE ISCAS. He served as AEiC for Digital Communications/Lead Guest Editor/Guest Editor/SEB Member for IEEE JETCAS, Associate Editor/Special Session Organizer for IEEE TCSVT, and Guest Editor for IEEE TCAS-II. He was Distinguished Lecturer of APSIPA and is Chair of IEEE CASS VSPC Technical Committee.
\end{IEEEbiographynophoto}
\vspace{-2em}
\begin{IEEEbiographynophoto}{Hsueh-Ming Hang} (M'78--SM'91--F'02) received the B.S. and M.S. degrees in control engineering and electronics engineering from National Chiao Tung University, Hsinchu, Taiwan, in 1978 and 1980, respectively, and Ph.D. in electrical engineering from Rensselaer Polytechnic Institute, Troy, NY, in 1984.

From 1984 to 1991, he was with AT\&T Bell Laboratories, Holmdel, NJ, and then he joined the Electronics Engineering Department of National Chiao Tung University (NCTU), Hsinchu, Taiwan, in December 1991. From 2006 to 2009, he took a leave from NCTU and was appointed as Dean of the EECS College at National Taipei University of Technology (NTUT). From 2014 to 2017, he served as the Dean of the ECE College at NCTU. He was appointed as the Dean of Faculty of System Engineering, NCTU in 2019. He has been actively involved in the international MPEG standards since 1984 and his current research interests include multimedia compression, multiview image/video processing, and deep-learning based image/video processing.

Dr. Hang holds 14 patents (Taiwan, US and Japan) and has published over 200 technical papers related to image compression, signal processing, and video codec architecture. He was an associate editor (AE) of the IEEE Transactions on Image Processing (1992-1994, 2008-2012) and the IEEE Transactions on Circuits and Systems for Video Technology (1997-1999). He is a co-editor and contributor of the Handbook of Visual Communications published by Academic Press in 1995. He was an IEEE Circuits and Systems Society Distinguished Lecturer (2014-2015).and is currently a Board Member of the Asia-Pacific Signal and Information Processing Association (APSIPA). He received the Distinguished Engineering Professor Award from Chinese Institute of Engineers (2005) and Chinese Institute of Electrical Engineering (2012). He is a recipient of the IEEE Third Millennium Medal and is a Fellow of IEEE and IET and a member of Sigma Xi.
\end{IEEEbiographynophoto}
\vspace{-2em}
\begin{IEEEbiographynophoto}{Marek Domański} received the M.Sc., Ph.D., and Habilitation degrees from the Poznań University of Technology, Poland, in 1978, 1983, and 1990, respectively. 

Since 1993, he has been a Professor with the Poznań University of Technology, where he leads the Institute of Multimedia Telecommunications. He coauthored one of the very first AVC decoders for tv set-top boxes, in 2004, highly ranked technology proposals to MPEG for scalable video compression, in 2004, 3-D video coding, in 2011, and immersive video coding, in 2019. He authored three books and more than 300 articles in journals and conference proceedings. His contributions were mostly on image, video and audio compression, virtual navigation, free-viewpoint television, image processing, multimedia systems, 3-D video and color image technology, digital filters, and multidimensional signal processing. 

Dr. Domański served as a member of various steering, program, and editorial committees of international journals and international conferences. He was the General Chairman/the Co-Chairman and the Host of several international conferences, including Picture Coding Symposium, PCS 2012; IEEE International Conference on Advanced Vedio and Signal Based Surveillance, AVSS 2013; European Signal Processing Conference, EUSIPCO 2007; 73rd and 112nd Meetings of MPEG; International Workshop on Signals, Systems and Image Processing, IWSSIP 1997 and 2004; and International Conference Signals and Electronic Systems, ICSES 2004.
\end{IEEEbiographynophoto}

\end{document}